\documentclass[11pt,letterpage(8.5in*11in)]{article}
\usepackage[utf8]{inputenc}

\usepackage{CJK}

\usepackage{amsmath,amsfonts,amssymb}

\usepackage{mathrsfs}

\usepackage{bm}

\usepackage{geometry}

\usepackage{ntheorem}

\usepackage{hyperref}

\usepackage{graphicx, subfig}

\usepackage{caption}

\usepackage[noend]{algpseudocode}

\usepackage{algorithmicx,algorithm}

\usepackage{enumerate}

\usepackage{setspace}

\newtheorem{theorem}{Theorem}[section]
\newtheorem{lemma}{Lemma}[section]
\newtheorem{claim}{Claim}
\newtheorem{definition}{Definition}[section]
\newtheorem*{hypothesis}{\it \#ETH:}
\newcommand{\pf}{\vspace{0.0em} \noindent {\bf Proof\ }}

\title{The Exponential-Time Complexity of the complex weighted \#CSP}
\date{}
\author{Ying Liu\thanks{Affiliation: State Key Laboratory of Computer Science, Institute of Software Chinese Academy of Sciences, and University of Chinese Academy of Sciences, China. \newline \indent Supported by NSFC61932002 and NSFC 61872340. \newline \indent Email : liuy@ios.ac.cn}}

\begin{document}
	
	\begin{spacing}{1.2}
		
		\maketitle
		
		\bibliographystyle{plain}

		\begin{abstract}
			
			In this paper, I consider a fine-grained dichotomy of Boolean counting constraint satisfaction problem (\#CSP), under the exponential time hypothesis of counting version (\#ETH). Suppose $\mathscr{F}$ is a finite set of algebraic complex-valued functions defined on Boolean domain. When $\mathscr{F}$ is a subset of either two special function sets, I prove that \#CSP($\mathscr{F}$) is polynomial-time solvable, otherwise it can not be computed in sub-exponential time unless \#ETH fails. I also improve the result by proving the same dichotomy holds for \#CSP with bounded degree (every variable appears at most constant constraints), even for \#R$_3$-CSP.
			
			An important preparation before proving the result is to argue that pinning (two special unary functions $[1,0]$ and $[0,1]$ are used to reduce arity) can also keep the sub-exponential lower bound of a Boolean \#CSP problem. I discuss this issue by utilizing some common methods in proving \#P-hardness of counting problems. The proof illustrates the internal correlation among these commonly used methods. 
			    
			~  

		\end{abstract}
		
		\clearpage
		
		\section{Introduction}

	\quad\ A classical sub-field in complexity theory is to provide classification of counting problems according their computation difficulty. As an analogue of NP, L. Valiant \cite{CCP} defined the class \#P as the set of problems which can be computed by nondeterministic polynomial time Turing machines with outputting the number of accepting computations. It is naturally that \#P-hardness and \#P-completeness are defined. Valiant also demonstrated some \#P-complete counting problems in his seminar papers \cite{CCP,completeproblems}, permanent and vertex cover problem for example. Inspiring by it, more \#P-complete or \#P-hard problems have been explored \cite{phardproblems} and a series of dichotomies has been presented along with fascinating counting frameworks established. One interesting framework is the constraint satisfaction problem (\#CSP). 
	
	Let a function set $\mathscr{F}$ as $\{F:D^k \to \mathbb{C}, k\in\mathbb{N}\}$. Complex weighted \#CSP($\mathscr{F}$) is defined as follow, which is similar as partition functions.
	
	\emph{Input: A finite set of constraints on variables $x_1,x_2,...,x_n \in D$. Each of them belongs $\mathscr{F}$ and has the form $F(x_{i_1},x_{i_1},...,x_{i_k})$.}   
	
	\emph{Output: $\sum_{x_1,x_2,...,x_n\in D}\prod F(x_{i_1},x_{i_1},...,x_{i_k})$}
	
	\#CSP is a general framework which can represent many counting problems, like \#VC(Vertex Cover). However, there are still many problems which can not be described as an \#CSP problem like \#Matching \cite{Reflection}, since \#CSP can only describe problems which have local constraints.
	
	A more general framework is defined in \cite{Holant}, named Holant Problem, which is inspired by Holographic Algorithms by L. Valiant \cite{LeslieHolographic,ValiantAccidental}. Given $\mathscr{F}$, Holant($\mathscr{F}$) accepts a signature grid $\Omega=(G, \pi)$ as input. $G(V,E)$ is a graph, $\pi$ labels every $v$ in $V$ with a signature $f_v\in \mathscr{F}$, and labels incident edges at $v$ with input variables of $f_v$. $\sigma$ is an assignment of all edges to $D$. Holant($\mathscr{F}$) outputs $\sum_{\sigma: E \to D} \prod_{v\in V} f_v(\sigma|_{E(v)})$. $E(v)$ consists all adjacent edges of $v$ and $\sigma|_{E(v)}$ denotes the restriction of $\sigma$ to $E(v)$. Holant Problem can express all counting problems.
	
	This paper focus on such $\mathscr{F}$ with domain $D=\{0,1\}$, that is to say, only Boolean \#CSP and Boolean Holant Problem are considered.   
	
\subsection{Related work}
	
	\quad\ A series of dichotomies of \#CSP have been developed, which is about either the problem polynomial time solvable or \#P-hard. Such dichotomies are also called ``FP vs \#P-hard" type. Creignou and Hermann have proved that \#CSP($\mathscr{F}$) can be solved in polynomial time only if every function in $\mathscr{F}$ is affine when $\mathscr{F}=\{F:\{0,1\}^k\to \{0,1\}\}$ \cite{BooleanCSP}. The tractable condition is $\mathscr{F}\subseteq\{F$ | $F$ is pure-affine$\}$ or $\mathscr{F}\subseteq\{F$ | $F$ is product-type$\}$ when the value domain of $\mathscr{F}$ expands to $\mathbb{Q}$ \cite{weightedCSP,rationalCSP}. Cai ,Lu and Xia defined two tractable classes $\mathscr{A}$ and $\mathscr{P}$ of complex weighted Boolean \#CSP \cite{complexCSPandR3CSP}. Cai and Chen developed such dichotomy to a natural culmination. They generalized above results and found the tractable condition of complex weighted \#CSP with variables assigned in any finite domain \cite{complexCSP}. 
	
	Based on the results of complex weighted Boolean \#CSP, more tractable classes of Boolean Holant problems have been discovered further. The dichotomies, of complex Holant* \cite{complexHolant*}, non-negative Holant \cite{nonnegativeHolant}, real Holant$^c$ \cite{realHolantc}, complex Holant$^c$ \cite{complexHolantc} and Real Holant \cite{realHolant}, are gradually proposed.
	
	Besides of the two general framework, there are also interesting researches on some special counting problems, like the complexity analysis for \emph{six-vertex model} \cite{sixvertexmodel}, \emph{graph homomorphism} \cite{GH}. 
	
	\emph{Exponential Time Hypothesis} (ETH \cite{ETHKAST,SETH}) is a better-known and widely-believed hypothesis, which says the satisfiability of 3-CNF formulas cannot be
	decided in time $exp(o(n))$. The parameter $n$ is the number of variables, and it can be replaced by the number of constraints according Sparsification Lemma \cite{SETH}. This hypothesis is used to find more tight lower bound of NP-hard problems.
	
	Dell et al. relaxed it to the counting version (\#ETH) and obtained sub-exponential lower bound of Tutte polynomial and the permanent problem when assuming \#ETH holds \cite{ETHpermTutte}. Curticapean showed a $2^{o(n)}$ time algorithm for some concrete counting problems would violate \#ETH, with using the idea of block interpolation \cite{blockinterpolation}. \#ETH causes the interest of whether existed ``FP vs \#P-hard" dichotomies have the same tractable condition as their ``FP vs \#ETH-hard" variant. It is yes for Boolean weighted Boolean \#CSP \cite{ETHCSP} and counting graph homomorphism \cite{ETHquantumGH}.

\subsection{Main results and proof outlines}

	\quad\ Inspired by above researches, this paper concentrates on discussing the complexity classification of complex weighted Boolean \#CSP under \#ETH. A powerful instrument is called pinning when reducing arity of functions, which is necessary in the proof of dichotomies. Pinning is applied to forcing variables to be $0$ or $1$, which is verified to have no effect on the \#P-hardness of a problem. The first step is preparing this instrument under \#ETH before completing aimed dichotomy. 
	
	\begin{theorem}
		If there exist $\varepsilon>0, D >0$ such that \#CSP($\mathscr{F} \cup \{\delta_0, \delta_1\}$) have no $O(2^{\varepsilon n})$ time algorithm even every variable appears in no more than $D$ constraints, then \#CSP($\mathscr{F}$) have no $O(2^{\varepsilon' N})$ time algorithm even every variable appears in at most $D'$ constraints, for some $\varepsilon', D' >0$. $n$ and $N$ are the variables number.
	\end{theorem}

	The proof is provided by two methods. One is block interpolation with categorized discussed on the character of functions in $\mathscr{F}$.There are also mixed with addition and subtraction among functions in this method. The other is using a special instance as \emph{black-box} gadget to construct reduction. A degree of universality among these methods are presented in the proof. 
	
	Now the dichotomy can be discussed. The two tractable classes $\mathscr{A}$ and $\mathscr{P}$ are two sets of functions, in which every function maps $\{0,1\}^{arity}$ to complex number. $\mathscr{P}$ is the set of the signatures which can be expressed as the product of some unary functions, binary equality function and binary dis-equality function. $\mathscr{A}$ consists of all functions in which support set is affine and non-zero values have the form $\mathfrak{i}^{P(x_1,x_2,...,x_n)}$ with variables $x_1,x_2,...,x_n$, where $\mathfrak{i}=\sqrt{-1}$ and P is a homogeneous quadratic polynomial over $\mathbb{Z}$ with every cross term's coefficient even.    

	\begin{theorem}
		Suppose $\mathscr{F}$ is a class of functions mapping Boolean inputs to complex numbers. If $\mathscr{F} \subseteq \mathscr{A}$ or $\mathscr{P}$, there is a polynomial time algorithm to solve \#CSP($\mathscr{F}$). Otherwise, there exist $\varepsilon>0$ and $D\in \mathbb{N}$ such that \#CSP($\mathscr{F}$) can not be solved in time $O(2^{\varepsilon n})$ even if every variable is required appearing in no more than $D$ constraints, when \#ETH holds. $n$ is the number of variables. 
	\end{theorem}
	
	The proof outline is consistent with proving ``FP vs \#P-hard" dichotomy of complex weighted Boolean \#CSP \cite{complexCSPandR3CSP}. The tractable part is trivial, so hardness under \#ETH is the point. Starting from one binary function $H$, Lemma 4.1 states the \#ETH-hardness of \#CSP($\{H\}$) when $H\notin \mathscr{A}\cup\mathscr{P}$, with help from Lemma 4.2 and Lemma 4.3. Lemma 4.4 shows the \#ETH-harness of \#CSP($\{F\}$) if the support set of $F$ is not affine, which is essential for Lemma 4.5 and Lemma 4.6. Lemma 4.5 and Lemma 4.6 provide the processes of reducing arity of a function when it is not in $\mathscr{A}$ or $\mathscr{P}$. Combining with these lemmas, when given a complex-value Boolean functions set $\mathscr{F}$ with $\mathscr{F}\not\subseteq\mathscr{A}$ and $\mathscr{F}\not\subseteq\mathscr{P}$, we can establish the reduction from \#CSP($\{H\}$) to \#CSP($\mathscr{F}$). The reduction keeps transmission of \#ETH-hardness so a $2^{o(n)}$ time algorithm of \#CSP($\mathscr{F}$) violates \#ETH. The dichotomy holds even restricting every variable appearing in at most $D$ constraints, which is also called bounded degree \#CSP. 
	
	It can be further proved that the hardness condition still keeps for \#R$_3$-CSP.
	
	\begin{theorem}
		If \#ETH holds, $\mathscr{F}\not \subseteq \mathscr{A}$ and $\mathscr{F}\not \subseteq \mathscr{P}$, then there exists $\varepsilon>0$ such that \#R$_3$-CSP($F$) has no $O(2^{\varepsilon n})$ time algorithm. $n$ is the number of input variables.
	\end{theorem}

	For convenience, all \#CSP problems are transferred to equivalent Holant Problem expressions to analysis. Lemma 5.1 presents a reduction from bounded degree \#CSP($\mathscr{F}$) to \#R$_3$-CSP($\mathscr{F}\cup \{H\}$), $H$ is an non-degenerate binary function. The rest part of section 5 proves how \#R$_3$-CSP($\mathscr{F}\cup \{H\}$) reduce to \#R$_3$-CSP($\mathscr{F}$), by using local holographic reduction and gadget construction. All reductions can complete in sub-exponential time and involved instances keep scale expanding linearly, so \#R$_3$-CSP($\mathscr{F}$) has same complexity classification with bounded degree \#CSP($\mathscr{F}$) under \#ETH.

		\section{Preliminaries}
	
	\quad\ This section is divided to three parts. Subsection 2.1 introduces some basic definitions and notations about functions and counting problems.  Subsection 2.2 presents the important theoretical basis, \emph{Exponential Time Hypothesis} of counting version.  Subsection 2.3-2.5 introduce three general methods in establishing reduction which are necessary for this article. 

\subsection{Definitions and notations}
	
	\quad\ There are some basic concepts about functions. 
	A binary function $F: \{0,1\}^2\to \mathbb{C}$ can be written as a $2\times2$ matrix $\begin{pmatrix}F(0,0) & F(0,1) \\ F(1,0) & F(1,1)\end{pmatrix}$. An $k$-arity function $F$ is symmetric function only if $F(x_1,x_2,...,x_k)=F(x_{\pi(1)},x_{\pi(2)},...,x_{\pi(k)})$, for any permutation $\pi: [k]\to [k]$. It can be expressed as $[f_0,...,f_k]$ when every variable is $0$ or $1$, $f_j$ is the value of $F$ on input of Hamming weight $j$. $(=_k)$ denotes equality function $[1,0,...,1]$ with arity $k$ and $(\neq_2)$ is binary dis-equality function $[0,1,0]$. 
	
	Two special unary functions are $\delta_0=[1,0]$ and $\delta_1=[0,1]$, which can force variables to be $0$ or $1$. For a $k$ arity function $F$, $F^{x_j=c}=F^{x_j=c}(x_1,...,x_{j-1},x_{j+1},...,x_k)=F(x_1,...,x_{j-1},c,x_{j+1},...,x_k)=F(x_1,...,x_k)\delta_c(x_j)$, $c\in\{0,1\}$. The operator obtains $F^{x_j=c}$ from $F$ by adding constraint $\delta_c$ on $x_j$, called \emph{pinning}. Another called \emph{projection} constructs $F^{x_j=*}=F^{x_j=*}(x_1,...,x_{j-1},x_{j+1},...,x_k)=\sum_{x_j}F(x_1,...,x_k)$ from $F$. Both operators are useful methods in reducing arity of functions.
	
	What follows are introductions of three special classes of functions, which are defined in Boolean domain.
	
	 \begin{definition}
	 	$\mathscr{D}=\{[a_1,b_1]\otimes[a_2,b_2]\otimes...\otimes[a_k,b_k]$ $|$ $a_j,b_j\in \mathbb{C},  j\in\{1,2,...,k\}\}$ is a set of constraints which are equivalent to tensor product of some unary functions. $k\in \mathbb{N}$. 
	 \end{definition}

	$\mathscr{P}$ is a super-set of $\mathscr{D}$.
	
	\begin{definition}
		A function belongs to $\mathscr{P}$ if and only if it can be expressed as product of some unary functions, binary equality function $([1,0,1])$ and binary dis-equality function $([0,1,0])$. 
	\end{definition}	

	It is observed that any function in $\mathscr{D}$ is degenerate (its corresponding matrix is singular). A binary function in $\mathscr{P}$ is either degenerate or of form $\begin{pmatrix} x & 0 \\ 0 & y \end{pmatrix}$ or $\begin{pmatrix} 0 & x \\ y & 0 \end{pmatrix}$, $x,y \in \mathbb{C}$.
	
	Suppose $X$ denotes $k+1$ dimension column vector $(x_1,x_2,...,x_k,1)^T$ over the Boolean field,  and $A$ is a Boolean matrix. $\chi_{AX}$ is the symbol of the affine relation on $x_1,x_2,...,x_k$, whose value is $1$ when $AX=\vec{0}$ and $\chi_{AX}=0$ otherwise.  
	
	\begin{definition}
		$\mathscr{A}$ is the class of all functions which have the form $\chi_{AX}\mathfrak{i}^{P(X)}$, where $\mathfrak{i}=\sqrt{-1}$ and $P(X)$ is a homogeneous quadratic equation over $\mathbb{Z}$, with additional requirement that every cross term has even coefficient. 
	\end{definition}
 
	The addition and multiplication in $P$ are the usual operations in $\mathbb{Z}$. The addition can be computed mod 4 since $\mathfrak{i}^4=1$. Functions in $\mathscr{A}$ have nice property to help build polynomial time algorithm, presented by  in Theorem 4.1 of Cai's article\cite{complexCSPandR3CSP}.   
	
	Given a class $\mathscr{F}$, in which every function maps Boolean variables to complex number. There are some details about \#CSP and Holant Problem as follows, which are defined on $\mathscr{F}$.
	
	\begin{definition}[\#CSP]
		
		\#CSP($\mathscr{F}$) is defined as:
		
		Input: An instance $I(V,\mathcal{C})$, $\mathcal{C}$ is a finite set of constraints $f_C$ on variables $v_1,v_2,...,v_n$, which have the form $f_C(v_{C,1},v_{C,2},...,v_{C,k_C})$. $f_C\in\mathscr{F}$.   
		
		Output: $Z(I)=\sum_{v_1,v_2,...,v_n \in \{0,1\}}$ $\prod_{\mathcal{C}} f_C(v_{C,1},v_{C,2},...,v_{C,k_C})$.
		 
	\end{definition}

	The instance $I(V,\mathcal{C})$ has graphical representation.  
	
	If all constraints are binary, the instance can be expressed an graph $G(V,E)$ with every vertex denotes one variable and every edge denotes a binary constraint on two adjacent variables. For example, counting the vertex cover of a graph $G(V,E)$ equals to input $G$ to \#CSP($\{[0,1,1]\}$) with attaching one variable $x_{v}$ to every vertex $v\in V$ and $OR_2$ $([0,1,1])$ to every edge $e\in E$. 
	
	It is more complex when some high arity constraints exist. Now the instance of \#CSP($\mathscr{F}$) can be treated as a bipartite graph $G(V_L\cup V_R, E)$, with $v\in V_L$ is attached with variable $x_v$ and $u\in V_R$ is attached with constraints $f_u\in \mathscr{F}$. A edge $e(v,u)\in E$ denotes that $f_u$ constrains $x_v$.         
	
	$\Delta$ usually denotes the maximum number of constraints which a variable appears in. An instance is bounded degree when $\Delta$ of it is bounded. \#R$_D$-CSP($\mathscr{F}$) consists of all \#CSP($\mathscr{F}$) problems where every variable appears in at most $D$ constraints, in other words, $\Delta$ of all instances of \#${\rm R_D}$-CSP($\mathscr{F}$) are no more than $D$. 
	
	Holant Problem is similar.
	
	\begin{definition}[Holant]
		
		A Holant problem defined by $\mathscr{F}$ (\#$\mathscr{F}$) is:
		
		Input: A signature grid $\Omega=(G,\pi)$. $G(V,E)$ is an undirected graph, and $\pi: V\to \mathscr{F}$ labels each $v\in V$ with a function $f_v$. The incident edges of $v$ are attached with variables constrained by $f_v$ so arity($f_v$) equal to the degree of $v$.
		
		Output: $Holant_{\Omega}(G)=\sum_{\sigma: E\to \{0,1\}} \prod_{v\in V} f_v(\sigma|_{N(v)})$. 
		
		The output also is denoted by \#$G$.
	\end{definition}
	
	$\sigma$ is an assignment of all variables which are attached to edges, and $\sigma|_{N(v)}$ is the assignment of $\sigma$ restricting to the incident edges of $v$ $(N(v))$. $\pi$ is usually omitted and $G$ represents the signature grid for simplify, in which has be attached with variables and functions. 
	
	A bipartite Holant problem $\# \mathscr{H} | \mathscr{F}$ is the same defined, where input $G(V_L\cup V_R, E)$ is bipartite. Each vertex in $V_L$ corresponds to a function in $\mathscr{H}$ and each vertex in $V_R$ denotes a constraint in $\mathscr{F}$. It is obvious that \#CSP($\mathscr{F}$) is equivalent to the Holant problem $\#\{=_1,=_2,=_3,...\} | \mathscr{F}$.
	
\subsection{Exponential Time hypothesis}
	
	\quad\ \emph{Exponential Time Hypothesis} introduced by Impagliazzo, Paturi, and Zane \cite{ETHKAST,SETH}, states a lower bound of solving 3-SAT (decide whether a 3-CNF formula is satisfied or not). The more relaxed counting version of it is \#ETH, which assume \#3-SAT have no sub-exponential time algorithm.  
 		
	\begin{hypothesis}
		There is a constant $\varepsilon>0$ such that no deterministic algorithm can compute
		\#3-SAT in time $exp(\varepsilon \cdot n)$, where $n$ is the number of variables. 
	\end{hypothesis}
	
	The lower bound can be strengthened to $exp(\varepsilon \cdot m)$ by sparsification lemma \cite{SETH}, which states that an arbitrary $k$-CNF formula can be reduced to the disjunction of at most $2^{\epsilon n}$ sparse $k$-CNF formulas in $O(poly(n)2^{\epsilon n})$ time for all $\epsilon>0$, $m$ is the number of constraints. A formula is sparse when $m=O(n)$, which is same as bounded degree. 
	
	A problem is called \#ETH-hard when it does not have $O(2^{\varepsilon \cdot n})$ time algorithms under \#ETH for some $\varepsilon>0$. It can claim that three operators (projection, pinning and gadget construction) do not affect the hardness of a \#ETH-hard problem, and the feature keeps even restricting all input instances bounded degree. It is trivial that \#CSP($\{F^{x_i=*}\}$) still is \#ETH-hard even requiring every variable appears in no more than constant constraints, if bounded degree \#CSP($\{F\}$) has no $O(2^{\varepsilon n})$ time algorithm under \#ETH. Such feature of pinning would be proved in Section 3. So there only argues about gadget construction.  

	\begin{claim}
		Suppose any function $f$ in $\mathscr{F'}$ can be constant size gadgets constructed by functions in $\mathscr{F}$. If \#CSP($\mathscr{F}$) has no $O(2^{\varepsilon n})$ time algorithm even every variable appears in no more than $D$ constraints, for some $\varepsilon>0$ and $D\in \mathbb{N}$, then there exist $\varepsilon'>0$ and $D'\in \mathbb{N}$ such that \#CSP($\mathscr{F'}$) has no $O(2^{\varepsilon' N})$ time algorithm even every variable appears at most in $D'$ constraints. $n$ and $N$ are the number of variables.
	\end{claim}
	
	Suppose the maximum size gadget has at most $C_1$ variables and $C_2$ functions, with max-degree $d$. $C_1,C_2$ and $d$ are constants. Give an instance $G(V_{variables}\cup V_{functions},E)$ of bounded degree \#CSP($\mathscr{F_1}$) with $n$ variables and $m$ constraints ($m\leq Dn$). A bounded degree graph $G'(V'_{variables}\cup V'_{functions},E')$ is constructed by replace every vertex in $V_{functions}$ by corresponding gadget in $poly(m)$ time. $|V'_{variables}|\leq n+C_1m$ and $|V'_{functions}|\leq C_2m$. 
	
	Assume for all $\varepsilon'>0$ and $D'\in \mathbb{N}$, \#CSP($\mathscr{F'}$) has a $O(2^{\varepsilon' N})$ time algorithm even every variable appears at most in $D'$ constraints. Then $Z(G')$ can be computed in time $O(2^{\varepsilon'(n+C_1m)})$ by choosing appropriate $D'$. Hence, there is a $poly(m)+ O(2^{\varepsilon'(n+C_1m)})$ time algorithm to compute $Z(G)$, which means \#CSP($\mathscr{F}$) can be computed in $O(2^{\varepsilon n})$ time even bounded degree, with $\varepsilon'=\varepsilon$. It causes a contradiction, so the assumption is wrong.

\subsection{Interpolation}

	\quad\ Interpolation is a universal tool in establishing reductions between counting problems. Polynomial interpolation is widespread applied to prove ``FP vs \#P-hard" dichotomies, but it does not work when turning to ``FP vs \#ETH-hard". Fortunately, Curticapean introduces a new useful framework \cite{blockinterpolation}, dub block interpolation, which bases on multivariate polynomial interpolation. Firstly I introduce polynomial interpolation by reducing \#CSP($\{[0,1,1]\}$) to \#CSP($\{[0,1,x]\}$), where $x\in \mathbb{C}$ and $x$ is not a root of unity.
	
	Given an instance $G(V,E)$ of \#CSP($\{[0,1,1]\}$), $|V|=n$ and $|E|=m$, $Z(G)$ can be computed with the help of the oracle of \#CSP($\{[0,1,x]\}$) in polynomial time.  
	 
	\begin{enumerate}[\quad]
		\item \emph{ \textbf{Step 1: (Set up interpolation)} } \quad Suppose type $t$ denotes the number of constraints which accept $(1,1)$ in one satisfied assignment. Then $Z(G)=\sum_{t\in\{0,1,...,m\}}\rho_t$, $\rho_t$ is the number of satisfied assignments which have type $t$. For a indeterminate $y$, a polynomial $\mu(y)$ is defined via
		
		\begin{equation}
			\mu(y)=\sum_{t\in\{0,1,...,m\}}\rho_t y^t. 
		\end{equation}
		 All $\rho_t$ can be recovered when given the value of $\mu$ at $m+1$ distinct point, then $Z(G)$ is solved.
		
		\item\emph{ \textbf{Step 2: (Recover coefficients from polynomial equations) }} \quad $G_k$ is obtained by replacing every edge in $G$ by $k$ parallel edges, in which every edge attached with function $[0,1,x]$, $k\in\mathbb{N}$.  $Z(G_k)=\mu(x^k)$. Since $x$ is not the root of unity, we can obtain $\mu$ at $m+1$ distinct point by choosing different $k\in[m+1]$ and querying the oracle of \#CSP($\{[0,1,x]\}$) to get $Z(G_k)$. Then we can use Lagrange interpolation to recover all coefficients, $Z(G)$ can be computed.
	\end{enumerate}
	
	A  system of $(m+1)$ equations is established to solve all $\rho_t$ in $(m+1)(poly(m)+OT)$ time, by constructing $(m+1)$ different $G_k$ to query oracle. 
	Suppose $T_{oracle}(S)$\footnote{We use the same notation in this article, to denote the  time cost of oracles which we wanted. }  denotes the time cost of an oracle which can solve any instance of \#CSP($\{[0,1,x]\}$), where $S$ is the scale of input  graph, then total time about computing $Z(G)$ is $(m+1)(poly(m)+T_{oracle}(S))+poly(m+1)+poly(m+1)$, in which first $poly(m+1)$ is the time to solve the system and second is to add all $\rho_t$.
	Suppose bounded degree \#CSP($\{[0,1,1]\}$) is \#ETH-hard, it can only get the result that \#CSP($\{[0,1,x]\}$) can not be computed in $O(2^{\varepsilon'\sqrt{m}})$ for some $\varepsilon'>0$  by such reduction, since $S=O(m^2)$. The gap between lower bounds of the two counting problems is produced by the oversize of $G_k$, in which there are at most $m(m+1)$ constraints. The polynomial reduction also bring $G_k$ with maximum degree $(m+1)$, so the result can not be trivial improved to bounded degree \#CSP($\{[0,1,x]\}$).
	
	Block interpolation is introduced to solve the gap by constructing exponential number $G_k$ with keeping $O(m)$ size. Dividing $E$ to $\frac md$ blocks and replacing every edge in each block by different $O(d)$ size gadgets independently, a multivariate polynomial with $(d+1)^{\frac md}$ terms and interpolate all coefficients after obtaining the value of the polynomial at $(d+1)^{\frac md}$ distinct points. The total time of new reduction is $(d+1)^{\frac md}(poly(m)+T_{oracle}(O(m)))+poly((d+1)^{\frac md})+poly((d+1)^{\frac md})$. By the new reduction, we can prove that \#CSP($\{[0,1,x]\}$) have no $O(2^{\varepsilon'm})$ time algorithm when \#ETH holds. If the gadgets we used to replace every constraint is bounded degree, then \#CSP($\{[0,1,x]\}$) still is \#ETH-hard even restricting that every variable appears in no more $D$ constraints, $D$ is a constant.
		
	The above example shows the power of block interpolation in transferring \#ETH-hardness of bounded degree counting problems. It has been applied to prove the sub-exponential lower bound of some discrete problems, like unweighted permanent, counting the number of matching and of vertex cover sets in undirected graph \cite{blockinterpolation}. And it is also an important tool in the proof of ``FP vs \#ETH-hard" dichotomies, for example, about unweighted Boolean \#CSP\cite{ETHCSP} and (quantum) graph homomorphism\cite{ETHquantumGH}. Theorem 2.1 states the \#ETH-hardness of \#CSP($\{[0,1,1]\}$), which is start problem in the reduction chain of Theorem 1.2.
	                      
	\begin{theorem}
		(Curticapean \cite{blockinterpolation}) If \#ETH holds, then there exist $\varepsilon>0$ and $D\in \mathbb{N}$ such that counting the number of vertex cover sets of $n$-vertices undirected graph $G$ have no $O(2^{\varepsilon n})$ time algorithm, even $G$ is simple and of maximum degree $D$. 
	\end{theorem}

\subsection{Holographic reduction}
	
	\quad\ The above methods, including gadget construction and interpolation, both build new instance $G'$ by using certain gadgets to replace functions in origin instance $G$. Such methods map the solution fragments of $G$ one-to-one or one-to-many to $G'$'s. There is another reduction method \emph{holographic transformation}, introduced by \cite{LeslieHolographic}, which is many-to-many map. It is also feasible since the relation between final sums is the point rather than the relation between concrete solution fragments. 
	
	The following illustrates how holographic reduction works.
	
	For an instance $G$ of $\#\mathscr{F}|\mathscr{H}$, we can add two vertices to every edge, which are attached with the signature $T^{-1}$ and $T$, and divide each edge to a $3$ length path. $T$ is an invertible $2\time2$ matrix belong to $GL_2(\mathbb{C})$. The new graph $G'$ is equivalent to an instance of$\#\tilde{\mathscr{F}}|\tilde{\mathscr{H}}$ and $\#(G')=\#(G)$. $\tilde{\mathscr{F}}=\{F(T^{-1})^{\otimes arity(F)}| F\in \mathscr{F}\}$ and $\tilde{\mathscr{H}}=\{T^{\otimes arity(H)}H| H\in \mathscr{H}\}$, with treating $F$ and $H$ as row vector of dimension $2^{arity(F)}$ and column vector of dimension $2^{arity(H)}$. Above is a holographic transformation (reduction) defined by $T$. A holographic reduction defined on $T^{-1}$ can also reduce $\#\tilde{\mathscr{F}}|\tilde{\mathscr{H}}$ to $\#\mathscr{F}|\mathscr{H}$. So the two problems are equivalent, which is Valiant’s Holant Theorem \cite{LeslieHolographic} states about.
	
	\begin{theorem}
		(Holant Theorem) For an invertible $2\times2$ matrix $T\in GL_2$, $\#\mathscr{F}|\mathscr{H}$ is equivalent to $\#\tilde{\mathscr{F}}|\tilde{\mathscr{H}}$, where $\tilde{\mathscr{F}}=\{F(T^{-1})^{\otimes arity(F)}| F\in \mathscr{F}\}$ and $\tilde{\mathscr{H}}=\{T^{\otimes arity(H)}H| H\in \mathscr{H}\}$.
	\end{theorem} 
	         
	If $T$ is an orthogonal matrix, then $\#\{=_2\}|\mathscr{F}$ is equivalent to $\#\{=_2\}|\tilde{\mathscr{F}}$ since $([1,0,1])(T^{-1})^{\otimes 2}= ([1,0]^{\otimes2}+[0,1]^{\otimes2})(T^{-1})^{\otimes 2}=(([1,0]T^{-1})^{\otimes2}+([0,1]T^{-1})^{\otimes2})=[1,0,1]$.

\subsection{Signature decomposition}
	
	\quad\ During the reductions among Holant problems, sometimes we can only construct the gadget which realize a signature $F=f^{\otimes l}$ rather than the aimed function $f$, where $l\in \mathbb{N}$. Fortunately, $f$ is also available according the decomposition theorem introduced by Lin and Wang\cite{nonnegativeHolant}.   
	
	\begin{theorem}
		(Lin and Wang\cite{nonnegativeHolant}) For any function set $\mathscr{F}$ and function $f$, \#$(\mathscr{F}\cup \{f\})$ $\leq_T$ \#$(\mathscr{F}\cup \{f^{\otimes l}\})$, $l\in\mathbb{N}$.  
	\end{theorem}

	This reduction also works under \#ETH, shown in Theorem 2.4.  

	\begin{theorem}
		For any function set $\mathscr{F}$ and function $f$, if \#$(\mathscr{F}\cup \{f\})$ has no $O(2^{\varepsilon n}) $ time algorithm for some $\varepsilon>0$ even every variable appears in no more than $D$ functions, then there exist $\varepsilon'>0$ such that \#$(\mathscr{F}\cup \{f^{\otimes l}\})$ can not be solved in $O(2^{\varepsilon'N})$ even every variable is not constrained by more than $D'$ constraints. $l,D,D'\in\mathbb{N^{+}}$ and $n,N$ are the number of variables.  
	\end{theorem}

	\pf{
		We follow the proof of Theorem 2.3. Assume the arity of $f$ is $k$. 
		
		\begin{enumerate}[1.]
			\item If $l=1$, it is trivial.
			
			\item Suppose the corollary is correct when $l<d$.
			\begin{enumerate}[(1).]
				\item There exists an instance $I$ of \#$(\mathscr{F}\cup \{f\})$ such that $\#I\neq 0$ and $f$ appears $p=cd+r(c\geq 0, 0<r<d)$ times in $I$. Replacing the first $cd$ signatures by $(f^{\otimes d})^{\otimes c}$ and the last $r$ functions by $f^{\otimes d}$ with bring new variables $y_1,y_2,...,y_{k(d-r)}$, then we get a new function $I'$ which is equivalent to $(\#I)f^{\otimes (d-r)}(y_1,y_2,...,y_{k(d-r)})$. There is a reduction from \#$(\mathscr{F}\cup \{f^{\otimes (d-r)}\})$ to \#$(\mathscr{F}\cup \{f\})$. Such reduction is constructed by a constant gadget $I'$ since $I$ is finite and determined. Hence, the corollary is true according to the assumption and Claim 1.
				
				\item For any instance $I$ of \#$(\mathscr{F}\cup \{f\})$, either $\#I=0$ or $f$ appears $cd(c\in \mathbb{N})$ times in $I$. Then there is an polynomial algorithm to solve $I$ with querying the oracle of \#$(\mathscr{F}\cup \{f^{\otimes l}\})$. Checking the number of $f$ in $I$, if it is the multiple of $d$, replace them by $f^{\otimes l}$ and query oracle, otherwise output $0$. It is obvious that the complexity of the two problems is equivalent in this case.       
			\end{enumerate} 
		\end{enumerate}
		 		
	}

	\newcounter{s}
	\setcounter{s}{2}
	
	The proof is an induction on $l$, and it uses an interesting trick. It argues about the instances of \#$(\mathscr{F}\cup \{f\})$. The condition gives either a special instance as a black-box gadget to establish reduction or a directly polynomial time algorithm with oracle. The idea is also adapted to the proof of Theorem 1.1 (Method-\Roman{s}) in Section 3.  
		
		\section{Pinning}
	
	\quad\ Pinning is an important operation in reducing arity of functions, which is essential when analyzing the complexity of \#CSP($\mathscr{F}$). Lemma 8 in \cite{weightedCSP} has been showed that the two unary functions $\delta_0=[1,0]$ and $\delta_1=[1,0]$ do not affect the \#P-hardness of non-negative weighted Boolean \#CSP. Following the same proof,
	such result can be automatically generalized for \#ETH-hardness even of complex weighted \#CSP,  with the help of high arity equality functions (even $=_n$). However, our aim is to use pinning to establish reduction between bounded degree \#CSP under \#ETH. So we need to find a new proof for Theorem 1.1, with only some low arity equality functions' help. 
	
	\newcounter{c}
	\setcounter{c}{1}
	\newcounter{d}
	\setcounter{d}{2}
	
	The following give two ways of that and both methods (\Roman{c} \& \Roman{d}) keep the degree still be constant.  
	~\\

\pf{\textbf{of Theorem 1.1:}
	Suppose $I(V_L\cup V_R, E)$ is an instance of \#CSP($\mathscr{F} \cup \{\delta_0, \delta_1\}$) with maximum degree $D$. $V_L$ is the set of all vertices which are attached with variables $x_1, x_2,..., x_n$. In $V_R$ there are $m$ vertices attached with $\delta_0$ or $\delta_1$ and $M$ vertices attached with functions belong to $\mathscr{F}$. So $|V_R|= M + m $ and $|E|\leq Dn$. We use $V_0/V_1$ to denote the set of all variable-vertices which are adjacent with $\delta_0 /\delta_1$. Suppose there is an oracle to solve \#CSP($\mathscr{F}$) with time $T_{oracle}(N)$, $N$ is the number of variables in  instances. 
	
	\begin{enumerate}[\Roman{c}.]
		\item Focusing on the feature of functions, it can be claimed at least one function in $\mathscr{F}$ can interpolate $\delta_0$ and $\delta_1$.
		
		\begin{enumerate}[1.]
			\item There exist a Boolean function $F\in \mathscr{F}$ which is not symmetric about the domain. That is to say, $\exists \tau=\tau_1,\tau_2,...,\tau_k\in \{0,1\}^k$, $F(\tau)\neq F(\bar{\tau})$ \footnote{It is easily confused with the concept of symmetric functions, which has been introduced in the bottom of page 4}. $\bar{\tau}=1-\tau_1,1-\tau_2,...,1-\tau_k$. A basic block interpolation can be applied.
			
			\quad\ Dividing each of $V_0$ and $V_1$ to $\frac{m}{d}$ blocks such that $V_0 = V_{10} \cup V_{20} \cup ... \cup V_{{\frac{m}{d}}0}$ and $V_1 = V_{11} \cup V_{21} \cup ... \cup V_{{\frac{m}{d}}1}$, with keeping $|V_{i0} \cup V_{i1}|\leq d$ for all $i \in \{1,2,...,\frac{m}{d}\}$, $d$ is a constant which can be chosen. A new instance $I'$ of $\#CSP({\mathscr{F}})$ is constructed by Combining all variables in $V_{i0}/V_{i1}$ as $t_{i_0}/t_{i1}$ and removing all adjacent $\delta_0/\delta_1$, for every $i\in [\frac{m}{d}]$. Suppose $T=(T_0,T_1)\in{\{0,1\}}^{2\frac md}$ is a type in which $T_0/T_1$ records all the assignment of $t_{i0}/t_{i1}$. $T_0=(t_{10}, t_{20},...,t_{\frac{m}{d}0})^T$ and $T_1=(t_{11}, t_{21},...,t_{\frac{m}{d}1})^T$. Then  $Z(I)=\rho_{\vec{0},\vec{1}}$ and $Z(I')=\sum_{T} \rho_T$, $\rho_T=\sum_{\sigma(\vec{x})\in \{0,1\}^{N-m}}F$.
			
			\quad\ If all $\rho_T$ can be computed then $Z(I)$ is obtained. Next we construct a system of equations to compute all $\rho_T$. A series of instances $I_{\vec{y}}$ can be constructed by add different binary function $f_{y_i}(t_{i0},t_{i1})$, $\vec{y}=\{1,2,3,4\}^{\frac md}$. The four binary function are $f_1=[1,1,1]$, $f_2=[1,0,1]$, $f_3=\begin{pmatrix} F(\vec{0})&F(\tau)\\F(\bar{\tau})&F(\vec{1}) \end{pmatrix}$ and $f_4=\begin{pmatrix} F(\vec{0})&0\\0&F(\vec{1}) \end{pmatrix}$. $f_1$ is equivalent to no operation, $f_2$ is equivalent to replace two input variables by one variable, $f_3(t_{i0},t_{i_1})=F(t_{i{\tau_1}},t_{i{\tau_2}},...,t_{i{\tau_k}})$ and $f_4=f_2\cdot f_3$.
			
			\quad\ A system of $4^{\frac md}$ equations is established by query the values of different $I_{\vec{y}}$. Each equation has the form:
			
			\begin{equation}
				Z(I_{\vec{y}})=\sum_{T} \rho_T \prod_{i=1}^{\frac md} f_{y_i}(t_{i0},t_{i1}).
				\label{equ3.1}
			\end{equation} 
			
			The coefficient matrix of the system is $\begin{pmatrix}
				1&1&1&1\\1&0&0&1\\F(\vec{0})&F(\tau)&F(\bar{\tau})&F(\vec{1})\\F(\vec{0})&0&0&F(\vec{1})
			\end{pmatrix}^{\otimes {\frac md}}$. If $F(\vec{0})\neq F(\vec{1})$, then the coefficient matrix is invertible so all $\rho_T$ can be recovered. All instances, which are used to query oracle, have maximum degree no more than $D\cdot d$ with size $O(N)$. The total time is $4^{\frac md}(poly(M+m)+T_{oracle}(O(N)))+poly(4^{\frac md})$. 
		
			\quad\ If $F(\vec{0})= F(\vec{1})$, we can not recover all $\rho_T$. But actually the aim is only the value of $\rho_{\vec{0},\vec{1}}$. Lemma 5.1 in Fu's paper\cite{sixvertexmodel} proved that we can merge the $\rho_T$ which has the same coefficient to obtain new system of equations. New coefficient matrix is $\begin{pmatrix}
				1&1&1\\1&0&0\\F(\vec{0})&F(\tau)&F(\bar{\tau})\end{pmatrix}^{\otimes {\frac md}}$, and it is full rank. The new system can be solved to get some $\rho_T$ and some partial sum of $\rho_T$. It happens that $\rho_{\vec{0},\vec{1}}$ has not been merged then the value has been recovered.    
			Another solution is using $\begin{pmatrix} 1&-1&0&0 \\ 0&0&1&-1 \end{pmatrix}$ to do line transformation to this system. The new coefficient matrix is full rank and $\rho_{\vec{0},\vec{1}}$ is also happened to be solved. 
			
			\quad\ However, the situation that coefficient matrix is not full rank can be avoided. We do the addition or subtraction about the values of instances before block interpolation. Suppose $\vec{l}=(l_1,l_2,...,l_{\frac{m}{d}}) \in \{1,2\}^{\frac{m}{d}}$, $I_{\vec{l}}$ is constructed on $I'$ by adding $f_{l_i}(t_{i0},t_{i1})$ for all $i\in[\frac md]$. Then $I'= I_{\vec{0}}$, and
			\begin{equation}
				Z(I'| \forall i=1,2,...,\frac{m}{d}, t_{i0}\neq t_{i1})= [1,-1]^{\otimes \frac md}\sum_{\vec{l}} Z(I_{\vec{l}})=\sum_{T|\forall i\in\{1,2,...,\frac{m}{d}\}, t_{i0}\neq t_{i1}} \rho_T.
				\label{equ3.2}
			\end{equation}  
		
			The left value can be computed in poly($2^{\frac md}$)-time by querying $2^{\frac md}$ times oracle of \#CSP\{$\mathscr{F}$\}. Based on Equation (3), we do block interpolation. Suppose $\vec{h}=h_1,h_2,...,h_{\frac md}\in \{0,1\}^{\frac md}$, adding $F(t_{i\tau_1},t_{i\tau_2},...,t_{i\tau_k})$ to obtain $I_{\vec{l},\vec{h}}$ when $h_i=1$. A system is established in which each equation has the form:
			\begin{equation}
				[1,-1]^{\otimes \frac md}\sum_{\vec{l}} Z(I_{\vec{l},\vec{h}})=\sum_{T|\forall i\in\{1,2,...,\frac{m}{d}\}, t_{i0}\neq t_{i1}} \rho_T \prod_{j=1}^{\frac md}{(F(\tau)^{1+t_{j0}}F(\bar{\tau})^{t_{j0}})}^{h_j}.
				\label{equ3.3}
			\end{equation} 
		
			There are $2^{\frac md}$ unknowns and the system size is also $2^{\frac md}$. According to the invertible coefficient matrix $\begin{pmatrix}1&1\\F(\tau)&F(\bar{\tau})\end{pmatrix}^{\otimes \frac md}$, we can solve $\rho_{\vec{0},\vec{1}}$. The time of building and solving the system is $2^{\frac md}[2^{\frac md}(poly(M+m)+T_{oracle}(N))+poly(2^{\frac md})]+poly(2^{\frac md})$. The max-degree of all instances is no more than $D\cdot d$ and all size are $O(n)$.
			
			\item For all function $F\in \mathscr{F}$, $F(\tau)= F(\bar{\tau})$ for all $\tau=\tau_1,\tau_2,...,\tau_k\in \{0,1\}^k$. Suppose $F'$ is obtained from $F$ by pinning. If any $F'$ is also symmetric, which means $F'(x)=F'(\bar{x})$ for any input, then $F$ is a constant-value function. \#CSP($\mathscr{F}$) is solvable in polynomial time. It contradicts the assumption in Theorem 1.1.
			
			\quad\ So for some $F'$, there must exist some $\tau=\tau_1,\tau_2,...,\tau_k \in \{0,1\}^{arity(F')}$ s.t. $F'(\tau)\neq F'(\bar{\tau})$. Suppose $arity(F')=k', arity(F)=k$ and $F'$ is obtained by pinning $x_{k'+1},...,x_k$ of $F$, then $F'= F^{x_j=0/1,j=(k'+1),...,k}$. If $\delta$ records the pinning assignments s.t. $F'(\tau)=F(\tau,\delta)$ and $F'(\bar{\tau})=F'(\bar{\tau},\delta)$. Reply on the feature, we can solve $\rho_{\vec{0},\vec{1}}$ by block interpolation.
			
			\quad\ Following case 1, there are only different when extending $I_{\vec{l}}$  to $I_{\vec{l},\vec{h}}$ : for all $i\in \{1,2,...,\frac md -1\}$, adding $F$ according $\tau$ to $I_{\vec{l}}$  when $h_i=1$ like Figure 1 showed.
			
			\begin{figure}[h]
				\centering
				\includegraphics[scale=0.5]{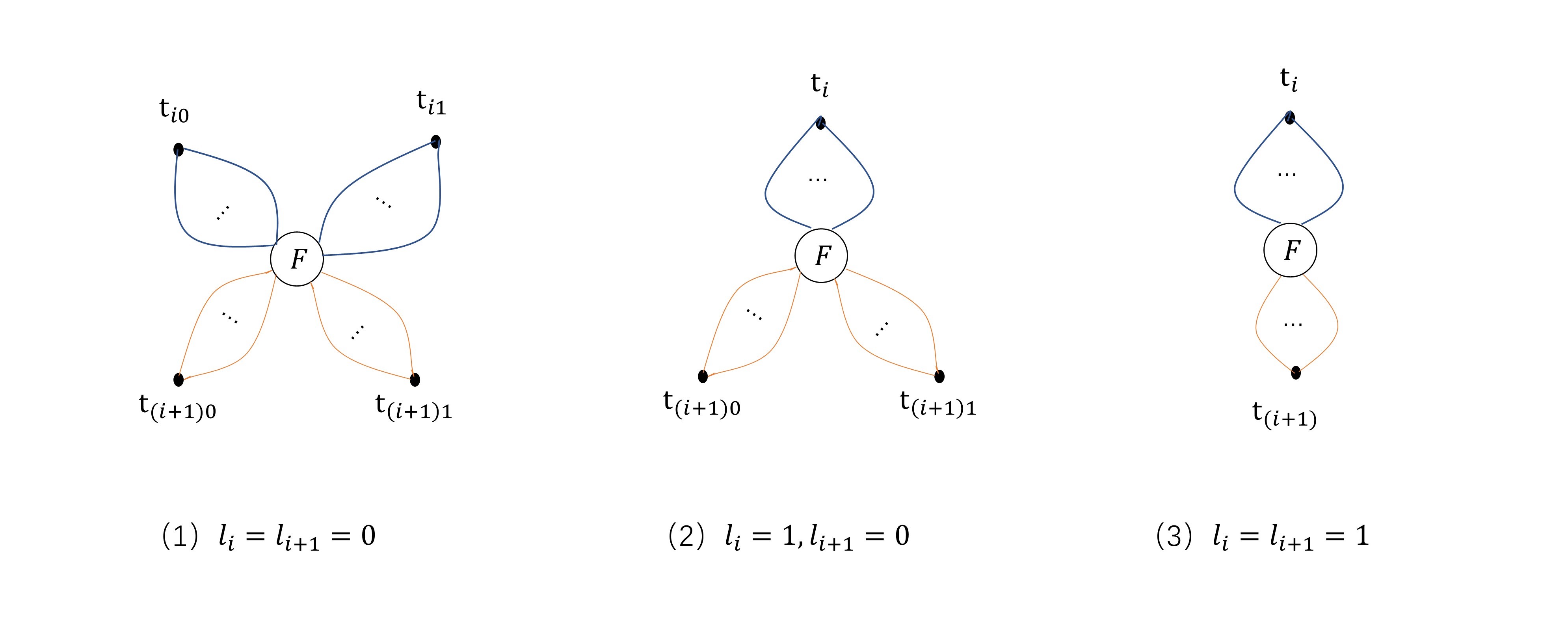}   
				\caption{ The blue edges mean the $1\sim k'$ inputs and the red are others. (1) $l_i = l_{i+1} = 0$, there are four variables $t_{i0},t_{i1},t_{(i+1)0}$ and $t_{(i+1)1}$. Considering the $j$-th input of $F$, if it should be pined to $0/1$, $t_{(i+1)0}/t_{(i+1)1}$ replaces it, otherwise $t_{i0}/t_{i1}$ replace it when $\tau_j=0/1$. (2) $l_i=1,l_{i+1}=0$, it is similar as (1) but $t_{i0}$ and $t_{i1}$ are merged to $t_i$. (3) $l_i=1,l_{i+1}=0$, we add $F(t_i,...,t_i,t_{i+1},...,t_{i+1})$, the first $k'$ inputs of $F$ are $t_i$}. 
			\end{figure}
			
			The addition of $F$ would bring the coefficient matrix:
			$\begin{pmatrix}
				F(\vec{0},\vec{0})&F(\vec{0},\delta)&F(\vec{0},\bar{\delta})&F(\vec{0},\vec{1})\\F(\tau,\vec{0})&F(\tau,\delta)&F(\tau,\bar{\delta})&F(\tau,\vec{1})\\F(\bar{\tau},\vec{0})&F(\bar{\tau},\delta)&F(\bar{\tau},\bar{\delta})&F(\bar{\tau},\vec{1})\\F(\vec{1},\vec{0})&F(\vec{1},\delta)&F(\vec{1},\bar{\delta})&F(\vec{1},\vec{1})
			\end{pmatrix}$. The row index is $t_{i0}t_{i1}$ and column index is $t_{(i+1)0}t_{(i+1)1}$. When only considering about $t_{i0}\neq t_{i1}$ and $t_{(i+1)0}\neq t_{(i+1)1}$, it transfers to a $2\times 2$ matrix:
			$\begin{pmatrix}
				F(\tau,\delta)&F(\tau,\bar{\delta})\\F(\bar{\tau},\delta)&F(\bar{\tau},\bar{\delta})
			\end{pmatrix}$. Because $F$ is symmetric, so $F(\tau,\delta)=F(\bar{\tau},\bar{\delta})$ and $F(\tau,\bar{\delta})=F(\bar{\tau},\delta)$. Suppose $F(\tau,\delta)=a$ and $F(\bar{\tau},\delta)=b$, then $a,b\in\mathbb{C}$ and $a\neq b$.
			
			When fix $\vec{h}$, we can construct such equation:
			\begin{equation}
				[1,-1]^{\otimes \frac md}\sum_{\vec{l}} Z(I_{\vec{l},\vec{h}})=\sum_{T|\forall i\in\{1,2,...,\frac{m}{d}\}, t_{i0}\neq t_{i1}} \rho_T \prod_{j=1}^{\frac md-1}{(a^{1+t_{j0}+t_{(j+1)0}}b^{t_{j0}+t_{(j+1)0}})}^{h_j}.
				\label{equ3.4}
			\end{equation}
			The plus between $t_{j0}$ and $t_{(j+1)0}$ is modulo-2 addition.
			
			\quad\ There are $2^{\frac md}$ unknowns in Equation \eqref{equ3.4}. Observing $\rho_T$ and $\rho_{\bar{T}}$ have same coefficient, we merge them as new unknown. So There are $2^{(\frac md-1)}$ unknowns. We can build equations system, which has $2^{(\frac md-1)}$ equations, by taking different $\vec{h}$. The coefficient matrix of the system is:   
			$\begin{pmatrix}
				1&1\\a&b
			\end{pmatrix}^{\otimes (\frac md -1)}$.
			The row index of the $2\times 2$ matrix is $h_j$ and The column index is $(t_{j0}+t_{(j+1)0})$ $mod$ $2$, $j=1,2,...,\frac md$. The coefficient matrix is invertible, since $a \neq b$, and we can compute all $\rho_T+\rho_{\bar{T}}$. Final we get $Z(I)=\rho_{\vec{0},\vec{1}}=\frac 12 (\rho_{\vec{0},\vec{1}}+\rho_{\vec{1},\vec{0}}))$.
			
			\quad\ In the process of solving $\rho_{\vec{0},\vec{1}}$, there are $2^{(\frac md -1)}$ equations and we need construct $2^{\frac md}$  instances to query oracle of \#CSP($\mathscr{F}$) to get an equation. It costs $2^{(\frac md -1)}[2^{\frac md}(poly(M+m)+T_{oracle}(N))+poly(2^{\frac md})]+poly(2^{(\frac md-1)})$ time together. The max-degree of all $I_{\vec{1},\vec{h}}$ are smaller than $D\cdot d$ and the size are $O(n)$.
			
		\end{enumerate} 
		
		\item[\Roman{d}.] The second method regards some instances as ``black-box" functions to help establish reduction, which is an effective tool in article \cite{complexCSP},\cite{nonnegativeHolant} and \cite{realHolant}. Suppose \#CSP$_q$($\{\delta_0\}\cup\mathscr{F}$) is the subset of \#CSP($\{\delta_0\}\cup\mathscr{F}$), in which all the instances have no more than $q$ variables attached with $\delta_0$. 
		Following is the reduction from \#CSP$_{\frac md}$($\{\delta_0\}\cup\mathscr{F}$) to \#CSP($\mathscr{F}$). 
		
		\quad\ Thinking about \#CSP$_{q+1}$($\{\delta_0\}\cup\mathscr{F}$) $\leq$ \#CSP$_q$($\{\delta_0\}\cup\mathscr{F}$). $I_{q+1}$ is an instance of \#CSP$_{q+1}$($\{\delta_0\}\cup\mathscr{F}$), showed in Figure 2. Removing the first $\delta_0$ to get an instance $I_q$ and treating $I_q$ as a unary function $I_q(x_1)$.
		
		\begin{enumerate}[(1).]
			\item If all $I_q(x_1)=[c, c]$, which means $I_q(0)$ always equals to $I_q(1)$, then $Z(I_{q+1})=I_q(0)={\frac 12}Z(I_q)$ since $Z(I_q)=I_q(0)+I_q(1)$.
			
			\item If there exists $I_q(x_1)=[a, b]$ with $a\neq b$, dub $f$, then $\delta_0$ can be obtained from $f=[a, b]$ as:
			$\delta_0=[1,0]=\frac 1{a-b}([a,b]-b[1,1])$. 
			This is equivalent to construct an instance $I_q^f$ by adding $f(x_1)$ to $I_q$, $Z(I_{q+1})=\frac 1{a-b}(Z(I_q^f)-bZ(I_q))$. $I_q^f$ and $I_q$ both are instances of \#CSP$_q(\{\delta_0\}\cup\mathscr{F})$. 
			
		\end{enumerate}
	
		\begin{figure}[h]
			\centering
			\includegraphics[scale=0.4]{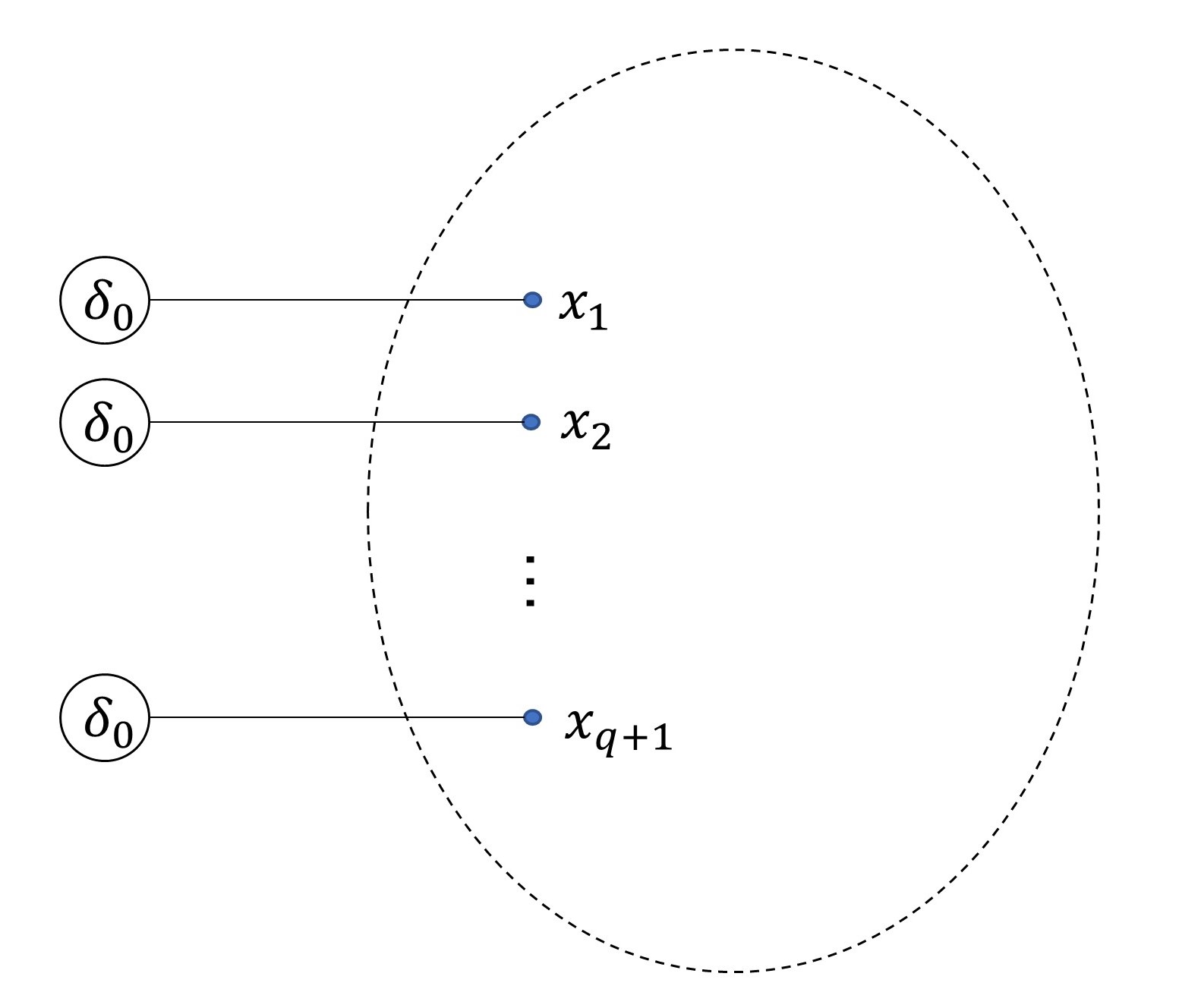}   
			\caption{There are $q$ variables $\{x_1,x_2,...,x_q\}$ adjacent with $\delta_0$, and the other variables and functions are invisible in the dotted circle.} 
		\end{figure}
		
		\quad\ To proof \#CSP$_{\frac md}(\{\delta_0\}\cup\mathscr{F})$ $\leq$ \#CSP$(\mathscr{F})$, we do \#CSP$_{\frac md}(\{\delta_0\}\cup\mathscr{F})$ $\leq$ \#CSP$_{\frac md-1}(\{\delta_0\}\cup\mathscr{F})$ $\leq...\leq$ \#CSP$(\mathscr{F})$. Such process can be expressed by an iterative tree of instances. The root is $I_{\frac md}$ and the edge is record the relationship between parent and child nodes like Figure 3 showed .The tree has no more than $\frac md$ depth and the degree of node at most 3. Totally, the oracle of $\#CSP(\mathscr{F})$ is queried no more than $2^{\frac md}$ times.
		
		\begin{figure}[ht]
			\centering
			\includegraphics[scale=0.4]{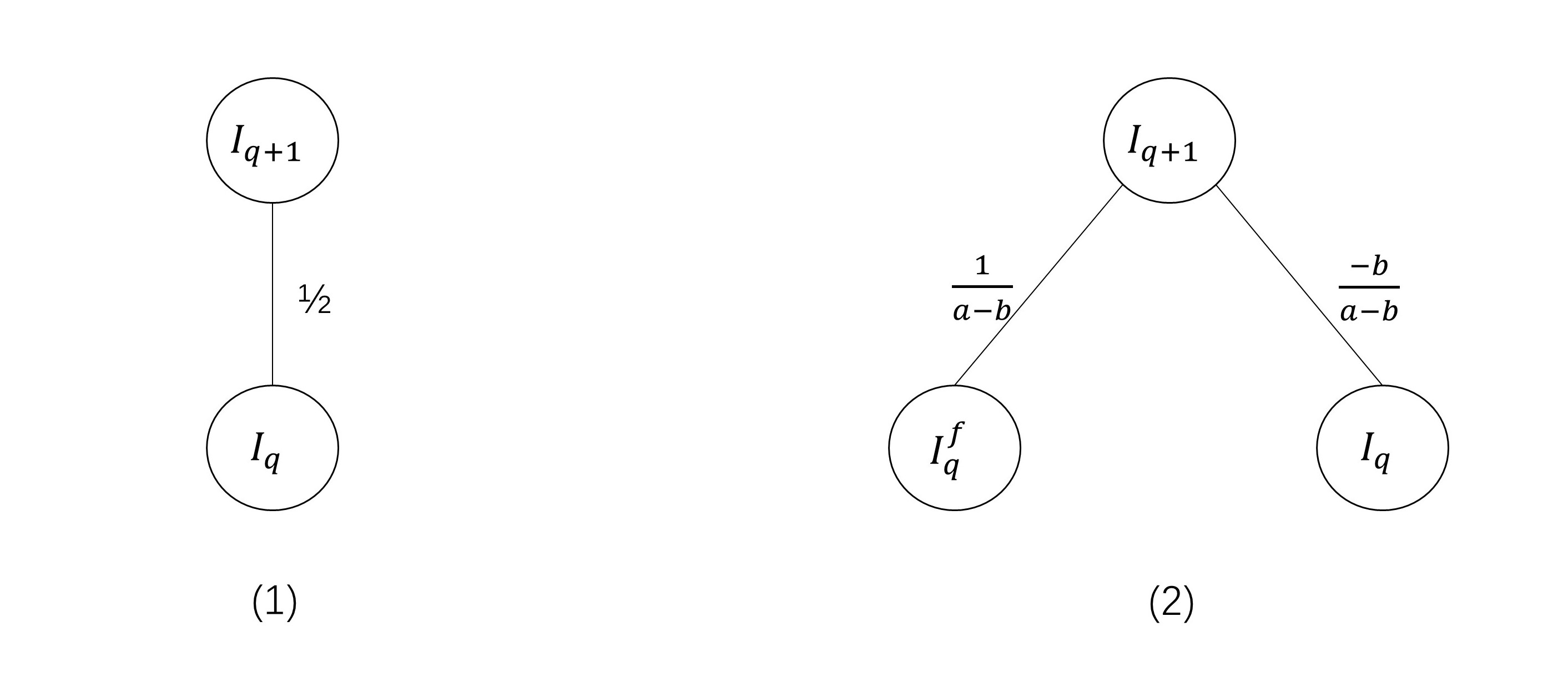}   
			\caption{(1) and (2) are corresponding to the different cases of $I_q$. The edge record coefficients of child instances when computing the value of parent instance.} 
		\end{figure}
		\quad\ Similarly, we construct $I'\in $\#CSP$_{\frac md}(\{\delta_0,\delta_1\}\cup \mathscr{F})$ like Method-I but reserving $\delta_0/\delta_1$. Then we construct iterative tree of  \#CSP$_{\frac md}(\{\delta_0,
		\delta_1\}\cup\mathscr{F})$ $\leq$ \#CSP$_{\frac md-1}(\{\delta_0,\delta_1\}\cup\mathscr{F})$ $\leq...\leq$ \#CSP$(\mathscr{F})$ in $poly(\frac md)$. Finally we compute $Z(I)=Z(I')$ according the tree by query $2^{\frac md}$ times oracle of \#CSP$(\mathscr{F})$. All costs $poly(\frac md)+2^{\frac md}(poly(M+m)+T_{oracle}(N))+poly(2^{\frac md})$ time. All involved instances have constant maximum degree and their sizes only change linearly.
	\end{enumerate}

	Concluding method I and \Roman{c}, there always exist reductions from \#CSP$(\{\delta_0,\delta_1\}\cup \mathscr{F})$ to \#CSP$(\mathscr{F})$ within $c\cdot(4^{\frac md}(poly(M+m)+T_{oracle}(N)))$ time, $c$ is a constant.
	
	Suppose for all $ \varepsilon'$, \#CSP($\mathscr{F}$) have a $c_12^{\varepsilon' N}$ time algorithm even bounded degree, $c_1$ is a constant. Then we can solve $I$ by querying a series of instances of \#CSP($\mathscr{F}$). The series of instances all have $N\leq c_2n$ variables and their maximum degree is smaller than $D'$, $c_2$ is also a constant. So the algorithm costs no more than $c_3 2^{2\frac {Dn}d +\varepsilon'c_1n}$, $c_3$ is an enough large constant which decided by $c$ and $c_2$. By choosing appropriate $d\ge \frac{4D}{\varepsilon}$ and $\varepsilon'\leq \frac{1}{2c_1}\varepsilon$, we get the algorithm of bounded degree \#CSP$(\{\delta_0,\delta_1\}\cup \mathscr{F})$ in $O(2^{\varepsilon n})$ time, which contradicts with original assumption. So the \#ETH-hardness of bounded degree \#CSP($\mathscr{F}$) has been proved.
}
	
	In method-I, the addition and subtraction among $I_{\vec{l}}$ actually is the process of using $f_1=[1,1,1]$ and $f_2=[0,1,0]$ to simulate the function $(=_2)$ as $[1,0,1]=[1,1,1]-[1,0,1]$. The addition and subtraction among instances sometimes correspond to addition and subtraction among functions. The addition and subtraction among existing functions is a method applied in proving the complexity of matching parameterized by genus\cite{parameterpermanent}.
	Such skill is also utilized to obtain $\delta_0/\delta_1$ in method-\Roman{c}. It can be seen all the methods in this proof (block interpolation, functions' addition or subtraction and the trick in method-\Roman{c}) have intercommunity. 

		\section{Dichotomy of complex weighted \#CSP}
	
	\quad\  Since the polynomial algorithms of \#CSP($\mathscr{A}$) and \#CSP($\mathscr{P}$) have been constructed in \cite{complexCSPandR3CSP}, this section focus on the hardness part of Theorem 1.2, which argues about that a \#P-hard \#CSP problem is also \#ETH-hard. Theorem 2.1 provides bounded degree \#CSP($\{[0,1,1]\}$) (\#VC) to establish reduction.  

\subsection{One binary function}

	\quad\ The start point is to prove the \#ETH-hardness of bounded degree \#CSP which only have a binary constraint.	
	
	\begin{lemma}
		If a binary function $H=\begin{pmatrix}a&b\\c&d\end{pmatrix}\notin \mathscr{A}\cup \mathscr{P}$, then there exist $\varepsilon>0$ and $D\in \mathbb{N}$ such that \#CSP(\{$H$\}) has no $O(2^{\varepsilon n})$ time algorithm under \#ETH, even every variable appears in at most $D$ functions. $n$ is the number of variables and $a,b,c,d\in \mathbb{C}$.
	\end{lemma}

	According to the definition, $H$ is non-degenerated since $H\notin \mathscr{P}$, and both $H$ and $H^T$ do not have the form like $\lambda[1,\pm \mathfrak{i},1],\lambda [1,\pm1,-1],\begin{pmatrix}1&-1\\\mathfrak{i}&\mathfrak{i}\end{pmatrix},\begin{pmatrix}1&-\mathfrak{i}\\\mathfrak{i}&-1\end{pmatrix},\begin{pmatrix}1&-\mathfrak{i}\\\-1&-\mathfrak{i}\end{pmatrix}$ because $H\notin \mathscr{A}$. To further simplify the problem, we restrict the value $a$ of $H$ to $0$.  
	
	\begin{lemma}
		If $H=\begin{pmatrix}0&b\\c&d\end{pmatrix}\notin \mathscr{A}\cup \mathscr{P}$, then bounded degree \#CSP($\{H\}$) is \#ETH-hard.
	\end{lemma}
	 
	\pf{\textbf{:}
		
		Since $H\notin \mathscr{A} \cup \mathscr{P}$, which means $bcd\neq 0$, we can normalize it and set b as $1$. Resetting $H$ as $\begin{pmatrix}0&1\\c&d\end{pmatrix}$. Suppose an instance of \#VC(which is equivalent to \#CSP($[0,1,1]$)) is a graph $G(V,E)$, in which $|V|=n,|E|=m$ and maximum degree is $\Delta$. The vertices represent variables and every edge is attached with $[0,1,1]$.
		
		Constructing $H'(x_1,x_2)=H(x_1,x_2)H(x_2,x_1)$, $H'=[0,c,d^2]=c\cdot[0,1,\frac{d^2}{c}]$. Resetting $H'$ as $[0,1,\frac{d^2}{c}]$. It can be claimed that \#VC is reduced to \#CSP($\{H'\}$).  
		
		\begin{enumerate}[1.]
		 
			\item $d'=\frac{d^2}{c}$ is a root of $1$.
			
			Suppose $(H')^k=[0,1,1]$, then constructing $G'\in$ \#CSP(\{$H'$\}) by replace every edge in $G$ by a $k$-length path, in which every edge represent the binary constraint $H'$. $Z(G)=Z(G')$.
			
			According Theorem 2.1, \#VC can not be solved in $O(2^{\varepsilon n})$ time for some $\varepsilon$.
			
			Suppose for all $\varepsilon'>0$, \#CSP({$H$}) have $O(2^{\varepsilon' N})$ time algorithm even bounded degree $2kD$. So $Z(G')$ is solved in $O(2^{\varepsilon' n'})$ time, $n'$ is the vertex number of $G'$. Then we can construct an algorithm to computing $Z(G)$ in time 
			$O(2^{\varepsilon n})$ by choosing $\varepsilon'=\varepsilon$. It contradicts to Theorem 2.1. So bounded degree \#CSP($\{H\}$) is \#ETH-hard.
			 
			\item $d'$ is not $1$'s root. 
			
			Block interpolation is utilized to establish reduction.
			For $G(V,E)$, dividing $E$ to $\frac md$ blocks with keeping every block has no more than $d$ edges, $d\in \mathbb{N}$. $E=B_1\cup B_2\cup ...\cup B_{\frac md}$ and $|B_i|\leq d$ for any $i\in\{1,2,...,\frac md\}$.    
			
			Suppose any vertex set $S\subseteq V$ has type $t=(t_1,t_2,...,t_{\frac md})^T\in \{0,1,2,...,d\}^{2\frac md}$ to record the number of edges covered by $S$. 		$t_i=(t_{i,1},t_{i,2})\in\{0,1,2,...,d\}$ and $t_{i,1}=|\{e|e\in B_i, |e\cap S|=1\}|,t_{i,2}=|\{e|e\in B_i, |e\cap S|=2\}|$. $x_t$ record the number of $S$ which has type $t$. $Z(G)=\sum_{t}x_t$. Suppose $\vec{y}=(y_1,y_2,...,y_{\frac md}) \in \mathbb{N}^{\frac md}$, we construct $G_{\vec{y}}(V,E')$ by replacing every edge in $B_i$ by $y_i$ parallel edges, which are attached with $H'$.
			\begin{equation}
				Z(G_{\vec{y}})=\sum_{t}x_t\prod_{i=1}^{\frac md}[(d')^{t_{i,2}}]^{y_i}.
				\label{equ4.1}
			\end{equation} 
			
			By choosing different $\vec{y}\in \{1,2,...,(d+1)^2\}^{\frac{m}{d}}$, we can get a system of equations which have form like \eqref{equ4.1}. In the system, there are exactly $(d+1)^{2\frac md}$ equations and unknowns. Noticing the coefficient matrix is Vandermonde matrix and invertible, so we can solve all $x_t$ in $poly((d+1)^{2\frac md})$ time. Then we plus all $x_t$ to obtain $Z(G)$. This algorithm costs $(d+1)^{2\frac md}\times(poly(m)+T_{oracle}(N_{G_{\vec{y}}}))+poly((d+1)^{2\frac md})+poly((d+1)^{2\frac md})=O(2^{m(\frac{log(d+1)}{d})}\times T_{oracle}(N_{G_{\vec{y}}}))$, $c$ is a constant. The first part is the time of building the system, by constructing $(d+1)^{2\frac md}$ times $G_{\vec{y}}$ to query the oracle of \#CSP(\{$H'$\}). The second and third part are the costing to solve all $x_t$ and add them. 
			
			Suppose for all $\varepsilon'>0$, \#CSP({$H$}) have $O(2^{\varepsilon' N})$ time algorithm even maximum degree is $2(d+1)D$. Then there is an algorithm to solve \#CSP({$H'$}) in time $O(2^{\varepsilon' n'})$ even the maximum degree of instances is $(d+1)D$, $n'$ is the number of input variables. By choose appropriate $d$ and $\varepsilon'$, which satisfy $D\frac{log(d+1)}{d}\leq \frac 12 \varepsilon$ and $\varepsilon'\leq \frac 12 \varepsilon$, $Z(G)$ can be computed in $O(2^{\varepsilon n})$. 
			It contradicts to Theorem 2.1, so bounded degree \#CSP($\{H\}$) is \#ETH-hard.
			
			More details are presented in Figure 4. 
		\end{enumerate}
		
		\begin{figure}[ht]
			\centering
			\includegraphics[scale=0.5]{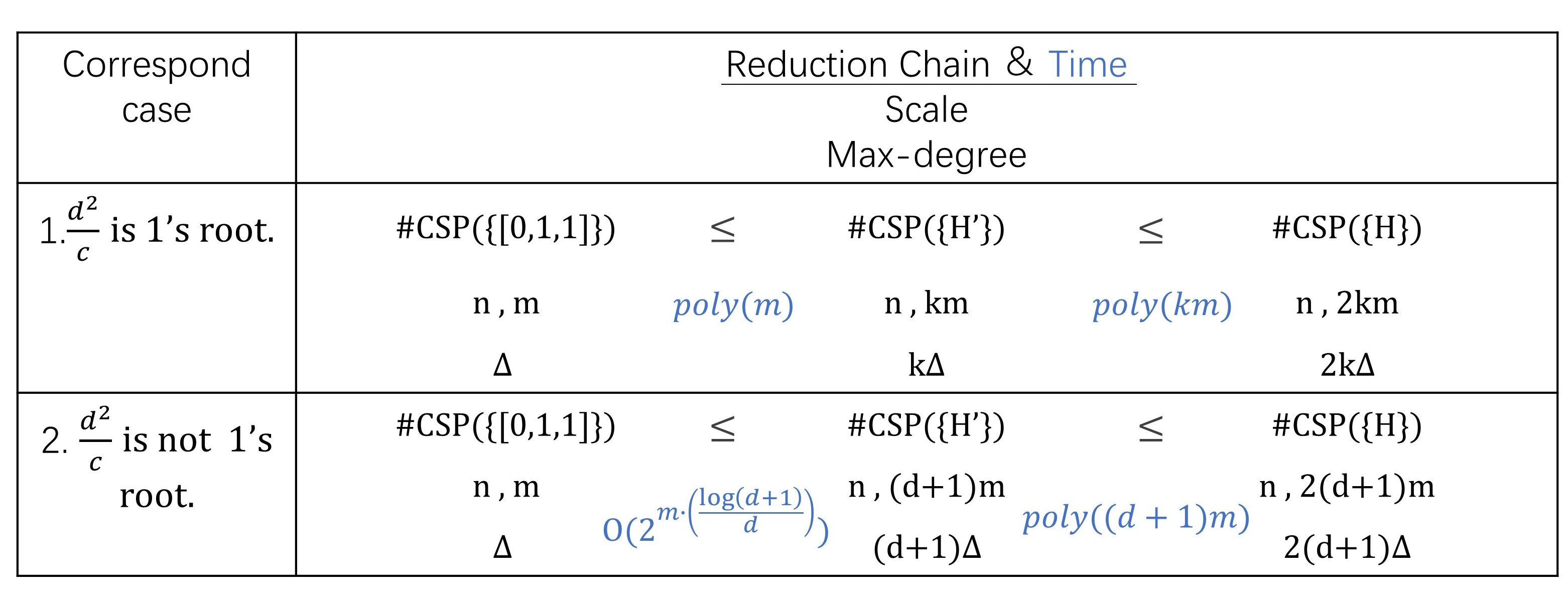}   
			\caption{ The maximum scale (variables number, functions number) and max-degree of instances in the reduction chains. Blue fonts are the time of reductions when assuming the oracle time is unity. k is a constant that $(H')^k=[0,1,1]$.} 
		\end{figure}
	}
	\begin{lemma}
			For a binary function $H=\begin{pmatrix}1&b\\c&d\end{pmatrix}$, $b,c,d\in \mathbb{C}$, if $d\neq bc$ and $bcd\neq0$ then there exist two unary function $[1,x]$ and $[1,y]$ such that $\#CSP(\{H,[1,x],[1,y]\})$ is \#ETH-hard even bounded degree.
	\end{lemma}

	This is because we can always construct a binary function not in $\mathscr{A}\cup\mathscr{P}$ by gadgets like the proof of Lemma 5.5 in \cite{complexCSPandR3CSP}). If $d\neq -bc$, We construct $H'(x_1,x_2)=\sum_{x_3}H(x_1,x_3)H(x_3,x_2)U_x(x_3)=\begin{pmatrix}0&{\frac{bc-d}{c}} \\ {\frac{bc-d}{b}}& {\frac{(bc)^2-d^2}{bc}}\end{pmatrix}$ by choosing $U_x=[1,-\frac 1{bc}]$. Otherwise, we construct $\begin{pmatrix}0&{-\frac 83 b} \\ {-\frac 83 c}& {\frac {80}{9} bc}\end{pmatrix}$ as
	
	\begin{equation}
		H''(x_1,x_2)=\sum_{x_3}(\sum_{x_4}H(x_1,x_4)H(x_4,x_3)U_x(x_4))(\sum_{x_5}H(x_3,x_5)H(x_5,x_2)U_x(x_5))U_y(x_3)
	\end{equation}
	by choosing $U_x=[1,-\frac 2{bc}]$ and $U_y=[1,-\frac 1{9bc}]$. The reduction from \#CSP(\{$H'$\}) or \#CSP(\{$H''$\}) to \#CSP(\{$H,U_x,U_y$\}) keeps the \#ETH-hardness by Claim 1, since the gadgets are constant size.
	
		With Lemma 4.2 and Lemma 4.3, we can complete the proof of Lemma 4.1.
	
	$\\$
	\textbf{Proof of Lemma 4.1:} 
	
	If $a=0$ ($d=0$ is symmetric), then the theorem has been proved by lemma 4.2. So we assume $a\neq 0$ and normalize $H=\begin{pmatrix}1&b\\c&d\end{pmatrix}$. Since $H \notin \mathscr{A}\cup \mathscr{P}$, then $bc\neq d, d\neq 0$ and at most one of the two value $b$ and $c$ can be $0$.
	
	\begin{enumerate}[1.]
		\item $bc\neq 0$,
		
		\begin{enumerate}[(1)]
			\item $d$ is not the root of $1$, then $[1,d]$ can be obtained from $H$. $[1,d]$ can interpolate all unary function like the part2 in proof of Lemma 4.1. Such block interpolations keep the translation of \#ETH-hardness. Since \#CSP(\{$H,[1,x],[1,y]$\}) is \#ETH-hard by Lemma 4.3, then \#CSP(\{$H$\}) is also \#ETH-hard. More details are presented in Figure 5.
			
			\item $d$ is the root of $1$ at least one of $\{b,c\}$ is not. Suppose $b$ is not $1$'s root.
			Then we can construct $U(x_1)=\sum_{x_2}H(x_1,x_2)\delta_0(x_2)=[1,b]$. Following case (1) to get \#ETH-hardness of \#CSP(\{$H,\delta_0$\}), then \#CSP(\{$H$\}) is also \#ETH-hard by Theorem 1.1.
			
			\item $b,c,d$ all are the root of $1$. Suppose $b^k=c^t=1$. By pinning one variable to $0$, we can get $U_1(x)=\sum_{x_1}H(x_1,x)\delta_0(x_1)=[1,b]=b^{k-1}[b,1]$ and $U_2(x)=\sum_{x_2}H(x,x_2)\delta_0(x_2)=[1,c]=c^{t-1}[c,1]$. Normalizing and 
			resetting $U_1(x)=[b,1]$ and $U_2(x)=[c,1]$. A symmetric function $G(x_1,x_2)$ is constructed by $G(x_1,x_2)=H(x_1,x_2)U_1(x_2)U_2(x_1)=bc[1,1,\frac d{bc}]$. $G\notin \mathscr{P}$ and $\frac d{bc}$ is a root of $1$.
			Considering whether $G$ is in $\mathscr{A}$.
			
			\begin{enumerate}
				\item[(a)] If $G\notin \mathscr{A}$, then $\frac{d}{bc}\neq -1$. We construct $H'=\sum_{x_2}G(x_1,x_2)=2[1,\frac{bc+d}{2bc}]$. It is easy to verify that $\frac{bc+d}{2bc}$ is not the root of $1$, so  all unary functions can be interpolated by $H'$. \#CSP(\{$G$\}) is \#ETH-hard by case (1). 
				
				\item[(b)] If $G\in \mathscr{A}$, then $\frac{d}{bc}= -1$ and $G=[1,1,-1]$. 
				
				\begin{enumerate}
					\item[(i)] $U_1=[1,b]$ or $U_2=[1,c]$ does not belong to $\mathscr{A}$. Suppose $U_1\notin \mathscr{A}$, then $b\notin \{\pm 1, \pm \mathfrak{i}\}$\footnote{We use $\mathfrak{i}$ to denote the imaginary unit with $\mathfrak{i}^2=-1$, and use $i$ to denote an integer index.}. We obtain $G'(x_1,x_2)=G(X_1,x_2)U_1(x_1)U_1(x_2)=2[1,b,-b^2]\notin \mathscr{A}\cup \mathscr{P}$ and further construct $H'=\sum_{x_2} G'(x_1,x_2)=(1+b)[1, \frac{b-b^2}{1+b}]$. Since $\frac{b-b^2}{1+b}$ is not the root of $1$, then it follows from case (1) that \#CSP(\{$G'$\}) is \#ETH-hard.  
					
					\item[(ii)] Both $U_1$ and $U_2$ belong to $\mathscr{A}$, then $b,c\in \{\pm 1, \pm \mathfrak{i}\}$. All satisfied $H$ are in $\mathscr{A}$, that contradicts with the assumption.
				\end{enumerate}
				
				If $G\in \mathscr{A}$, we can always construct a binary function $G'\notin \mathscr{A} \cup \mathscr{P}$ by gadgets of $G$. Naturally \#CSP(\{$G$\}) is \#ETH-hard since \#CSP(\{$G'$\}) is. 
			\end{enumerate}
		
			The above has proved the hardness of \#CSP(\{$G$\}), which can reduce to \#CSP(\{$H,\delta_0$\}) with keep \#ETH-hardness. Theorem 1.1 tells that \#CSP(\{$H$\}) is \#ETH-hard since pinning do not affect complexity. More details are shown in Figure 5.
		\end{enumerate}
		
		\item $b=0$(or $c=0$), then $H=\begin{pmatrix}1&0\\c&d\end{pmatrix}$. We consider $H'=\sum_{x_3}H(x_1,x_3)H(x_2,x_3)=[1,c,c^2+d^2]$, which is not in $\mathscr{P}$.
		
		If $H'\notin \mathscr{A}$, then \#CSP(\{$H$\}) is \#ETH-hard since \#CSP(\{$H'$\}) is by case 1.
		
		If $H'\in  \mathscr{A}$,  $H$ equals to either $\begin{pmatrix}1&0\\\pm \mathfrak{i}& \sqrt{2}\end{pmatrix}$ or $\begin{pmatrix}1&0\\\pm 1& \sqrt{2}\mathfrak{i}\end{pmatrix}$ since $H'$ has the form $[1,\pm \mathfrak{i},1]$ or $[1,\pm 1,-1]$. For all possible $H$, $d$ is not the root of 1. Case 1-(1) proves that \#CSP(\{$H$\}) is \#ETH-hard.
		
	\end{enumerate}
	
	Figure 5 shows the details of all reductions in this proof. The maximum degree of instances keep bounded in the translation of \#ETH-hardness, so \#CSP(\{$H$\}) has no $O(2^{\varepsilon n})$ time algorithm even the degree of input graphs is bounded.
	
	\begin{figure}[ht]
		\centering
		\includegraphics[scale=0.5]{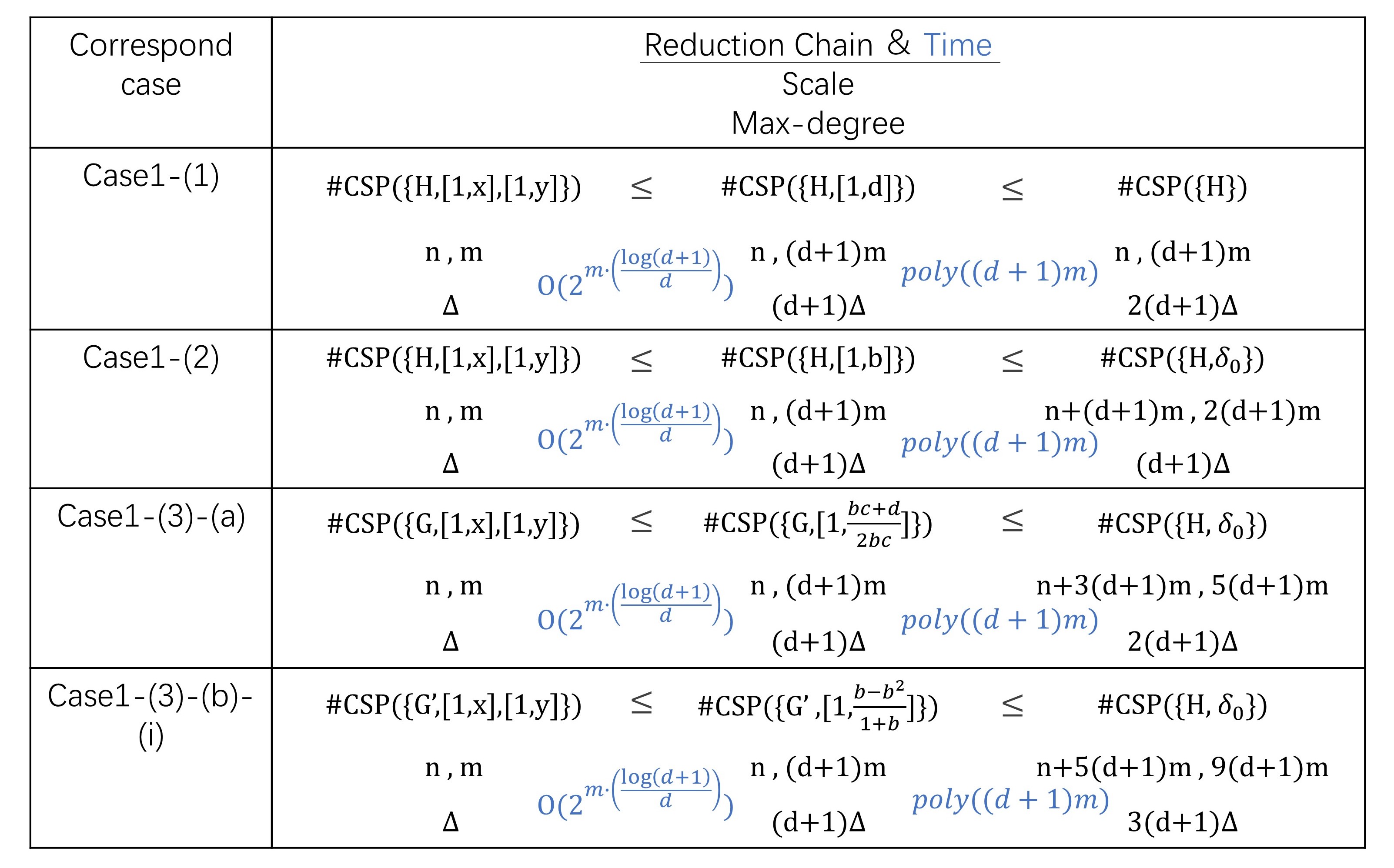}   
		\caption{The bound of instance's scale and max-degree in reduction chains of Lemma 4.1's proof. Blue fonts show the necessary time to complete these reductions with assuming the oracle time is unity.} 
	\end{figure}
	
\subsection{Reduce Arity}

	\quad\ Now we handle the situation when there is a function $F$ with high arity in a \#CSP problem. If $F$ does not have affine support, \#CSP($\{F\}$) is \#ETH-hard by Lemma 4.4. If $F\notin\mathscr{A}$ or $F\notin\mathscr{P}$, functions with smaller arity can be recursively simulated from $F$ by pinning or projection, keeping the property of being not in $\mathscr{A}$ or not in $\mathscr{P}$ respectively, according to Lemma 4.5 and Lemma 4.6.
	
	The support set $R_F$ of a function $F$ is variables' assignment set $\{\vec{x}|F(\vec{x})\neq0\}$, $\vec{x}\in \{0,1\}^{arity(F)}$. $R_F$ is affine if and only if $\vec{x_3}=\vec{x_1}\oplus\vec{x_2}$ is also in $R_F$ for any $\vec{x_1},\vec{x_2}\in R_f$, 
	
	\begin{lemma}
		Suppose $F$ is a function with arity $k$ and $R_F$ is not affine. If \#ETH holds, then there exist $\varepsilon>0$ and $D\in \mathbb{N}$ such that \#CSP(\{$F$\}) has no $O(2^{\varepsilon n})$ time algorithm, where n is the number of variables, even every variable appears in at most $D$ functions.  
	\end{lemma}
	
	\pf{
		We prove it by induction on function's arity. Because $F$ does not have affine support, so $k\geq2$.  
		\begin{enumerate}[1.]
		 \item $k=2$.
		 
		 \quad\ Suppose $F=\begin{pmatrix}a & b \\ c & d\end{pmatrix}$, exactly one of $a,b,c,d$ is zero since $R_F$ is not affine. Constructing $H=FF^T=[a^2+b^2,ac+bd,c^2+d^2]$, $H\notin \mathscr{P}$ and all elements of $H$ have non-zero value.
		 
		 \begin{enumerate}[(1).]
		 	\item $H\notin\mathscr{A}$. \#CSP(\{$H$\}) is \#ETH-hard even bounded degree by Lemma 4.1, so does \#CSP(\{$F$\}).
		 	
		 	\item $H\in\mathscr{A}$. $H$ has the form $x[1,\pm \mathfrak{i},1]$ or $x[1,\pm1,-1]$.
		 	\begin{enumerate}[(a)]
		 		\item $a=0$. $F$ is normalized to be $\begin{pmatrix}0 & 1 \\ c & d\end{pmatrix}$ and $H=[1,d,c^2+d^2]$.
		 		
		 		If $H=[1,\pm\mathfrak{i},1]$, then $F=\begin{pmatrix}0 & 1 \\ \pm \sqrt{2} & \pm \mathfrak{i}\end{pmatrix}$. Suppose $H'(x_1,x_2)=F(x_1,x_2)F(x_2,x_1)$, \#CSP(\{$H'$\}) is \#ETH-hard even bounded degree by lemma 4.2, since $H'\notin \mathscr{A}\cup\mathscr{P}$.
		 		
		 		It is similar when $H=[1,\pm1,-1]$.
		 		
		 		\item $b=0$. The \#ETH-hardness of \#CSP(\{$H$\}) can be verified by following case (a). It is different only when constructing $H'$. We obtain $H'$ by $H'=F^2$ here.  
		 	\end{enumerate}
		 \end{enumerate}  
		 
		 \item 
		 Suppose Lemma 4.4 holds for all functions with arity $k'< k$. The proof of Lemma 5.7 in \cite{complexCSPandR3CSP} shows we can always construct a smaller arity function $F'$ from $F$ by pinning or projection, with keeping $R_{F'}$ not affine. The reduction from \#CSP($\{F'\}$) to \#CSP($\{F\}$) keeps the \#ETH-hardness even the two problems are bounded degree, according to definitions and Theorem 1.1.
	\end{enumerate}

}

	Analyzing the proofs of Lemma 5.7 and Lemma 5.8 in \cite{complexCSPandR3CSP}, it is naturally to follow them since only gadget constructions are involved in the related reductions. Then we can claim the next two lemmas. 
	
	\begin{lemma}
		If $F\notin \mathscr{A}$, either \#CSP($\{F\}$) is \#ETH-hard or a unary function $H\notin\mathscr{A}$ can be simulated with the help of pinning or projection.
	\end{lemma}    
	
	If $R_F$ is not affine, \#CSP(\{$F$\}) is \#ETH-hard by Lemma 4.4. Focusing on the status that $R_f$ is affine, Cai \cite{complexCSPandR3CSP} provides the method to construct a unary function $H$ or a smaller arity function $F'$, with keeping both not in $\mathscr{A}$, by constant size gadgets of $F,\delta_0$ and $\delta_1$. Then a unary function $H\notin \mathscr{A}$ can be obtained recursively in polynomial time. And \#CSP($\{F\}$) would be \#ETH-hard if \#CSP($\{H\}$) is by Claim 1 and Theorem 1.1.
	
	\begin{lemma}
		
		If $F\notin \mathscr{P}$, either \#CSP($\{F\}$) is \#ETH-hard or we can simulate $[a,0,1,0]/[0,1,0,a]$ with $a\neq0$ or a binary function $H\notin \mathscr{P}$ having no zero value, with the help of pinning or projection. 
		
	\end{lemma}
	
	The proof is similar and related reductions are only established by gadget construction.
	
	\subsection{Proof of Theorem 1.2} 
	
	\quad\ There are polynomial algorithms for \#CSP($\{\mathscr{A}\}$) and \#CSP($\mathscr{P}$) in Cai's article\cite{complexCSPandR3CSP}. For \#CSP($\mathscr{P}$), any input graph can be divided into some connected components by replace each function by its factors, according the definition of $\mathscr{P}$. The variables in one connected component are constrained by $=_2$ and $\neq_2$, so there only two assignment for each component. The value of each component is easily computed and the value of the input graph is the product of its value on each connected component. For \#CSP($\mathscr{A}$), any instance's value can be computed recursively in polynomial time.   
	
	The point is considering the hardness. If $\mathscr{F}\not\subseteq \mathscr{A}$ and $\mathscr{F}\not\subseteq \mathscr{P}$, then there exist $f,g\in \mathscr{F}$ with $f\notin \mathscr{A}$ and $g\notin \mathscr{P}$. Considering \#CSP(\{$f,g$\}), either it is \#ETH-hard even bounded-degree by Lemma 4.4, or \#CSP(\{$F,P$\}) or \#CSP(\{$F,H$\}) can be reduced to it by Lemma 4.5 and Lemma 4.6. $F=[1,\lambda]\notin \mathscr{A}$, $P$ is $[a,0,1,0]([0,1,0,a])$ with $a\neq0$ and $H$ is a no-zero-value binary function which is not in $\mathscr{P}$. The reductions keep the transmission of \#ETH-hardness even all problems are bounded degree. So now the \#ETH-hardness of \#CSP(\{$F,P$\}) and \#CSP(\{$F,H$\}) is the aim even if both are bounded degree.
	
	A binary function $Q\notin \mathscr{A}\cup\mathscr{P}$, can be constructed by $\{F,H\}$ or $\{F,P\}$ respectively, with the help of pinning or projection. 
	
	\begin{enumerate}[1.]
		\item Constructing $P^{x_1=*}=[a,1,1]$. If it is not in $\mathscr{A}\cup\mathscr{P}$, then $Q=P^{x_1=*}$. Otherwise, $a=\pm1$, then constructing  $Q(x_1,x_2)=\sum_{x_3}P(x_1,x_2,x_3)F(x_3)=[\pm 1,\lambda,1]$. $\lambda$ is not a power of $\mathfrak{i}$ since $F\notin \mathscr{A}$. It can be verified that $Q\notin \mathscr{A}\cup\mathscr{P}$.
		
		\item Suppose $H=\begin{pmatrix} 1 & x \\ y & z \end{pmatrix}\notin\mathscr{P}$ with $xyz\neq 0$ and $z\neq xy$. If $H\notin \mathscr{A}$, $Q=H$. Otherwise $H\in \mathscr{A}$ then $z=-xy$. We can construct $Q(x_1,x_2)=\sum_{x_3}H(x_1,x_3)H(x_2,x_3)F^s(x_3)=(1+\lambda^sx^2)[1,\frac{y(1-\lambda^sx^2)}{1+\lambda^sx^2},y^2]$, which decides by $s\in \{0,1\}$. Because $\lambda$ is not the power of $\mathfrak{i}$, at most one of the two value $x^2$ and $\lambda x^2$ can be a power of $\mathfrak{i}$. $Q$ is not in $ \mathscr{A}\cup \mathscr{P}$ by choosing $s=0/1$ to force $\lambda^s x^2\notin \{\pm1,\pm \mathfrak{i}\}$. 
	\end{enumerate}  

	By Lemma 4.1, \#CSP(\{$Q$\}) is \#ETH-hard even bounded degree, then \#CSP($\mathscr{F}$) is also \#ETH-hard according the above analysis.

		\section{Dichotomy of \#R$_3$-CSP}

	\quad\ Based on Section 4, it has been known the dichotomy of \#CSP with bounded degree. This result holds even restricting the bound of degree to $3$. The tractability still applies of course. This section is aimed to prove the \#ETH-hardness of \#R$_3$-CSP($\mathscr{F}$) when $\mathscr{F}\not\subseteq\mathscr{A}$ and $\mathscr{F}\not\subseteq\mathscr{P}$. It is deserved to mention the following proofs also provide one way to prove Theorem 1.1, with additional using the property $\mathscr{F}\not\subseteq \mathscr{A}$ and  $\mathscr{F}\not\subseteq \mathscr{P}$.
	
	For convenience, all \#CSP problems are transferred to equivalent bipartite Holant problems to analyze. \#R$_3$-CSP($\mathscr{F}$) is equivalent to \#\{$=_1,=_2,=_3$\}|$\mathscr{F}$ and bounded degree \#CSP($\mathscr{F}$) is equivalent to \#\{$=_1,=_2,=_3,...,=_D$\}|$\mathscr{F}$. To build reduction from \#\{$=_1,=_2,=_3,...,=_D$\}|$\mathscr{F}$ to \#\{$=_1,=_2,=_3$\}|$\mathscr{F}$, first step is replacing any $=_k$ function by an equivalent tree gadget with $logk$ depth, whose root is attached with $=_2$, other nodes are attached with $=_3$ and every edge is put an extra node attached with $=_2$ to keep the bipartite. The equivalent construction has $(k-1)+2(k-1)$ vertices and $2(k-1)$ edges. Thus, a reduction is established from \#\{$=_1,=_2,=_3,...,=_D$\}|$\mathscr{F}$ to \#\{$=_1,=_2,=_3$\}|$\mathscr{F}\cup \{=_2\}$ by such tree gadgets. By Theorem 1.2 and Claim 1, \#\{$=_1,=_2,=_3$\}|$\mathscr{F}\cup \{=_2\}$ is \#ETH-hard if $\mathscr{F}\not\subseteq \mathscr{A}$ and  $\mathscr{F}\not\subseteq \mathscr{P}$.
	
	It turns to prove that \#\{$=_1,=_2,=_3$\}|$\mathscr{F}\cup \{=_2\}$ can be reduced to \#\{$=_1,=_2,=_3$\}|$\mathscr{F}$ with keeping \#ETH-hardness transmission. Importing a non-degenerate binary function $H$ and using \#\{$=_1,=_2,=_3$\}|$\mathscr{F}\cup \{H\}$ as intermediate problem. 
	
	\begin{lemma}
		Let $H$: $\{0,1\}^2 \to \mathbb{C}$ is a non-degenerate binary function. $\mathscr{F}$ is a set of complex value functions defined on Boolean domain, with $\mathscr{F}\not\subseteq\mathscr{A}$ and $\mathscr{F}\not\subseteq\mathscr{P}$. If \#ETH holds, then there exists $\varepsilon>0$ such that $\#\{=_1,=_2,=_3\}|\mathscr{F}\cup \{H\}$ has no $O(2^{\varepsilon n})$ time algorithm. $n$ is the number of functions in the left part of an instance.
	\end{lemma}
	
	\pf{\textbf{:}
		The reduction is established block interpolation.The Jordan normal of $H$ is more convenience when explaining the interpolation. $H$ can be decomposed as $T\Lambda T^{-1}$ for some invertible matrix $T$, in which $\Lambda$ has the form $\begin{pmatrix} \lambda_1 & 0 \\ 0 & \lambda_2\end{pmatrix}$ or $\begin{pmatrix} \lambda & 1 \\ 0 & \lambda \end{pmatrix}$. 
		
		Given a graph $G(V_l\cup V_r, E)$ as an instance of $\#\{=_1,=_2,=_3\}|\mathscr{F}\cup \{=_2\}$. $|V_l|=N$, $|V_r|=M\leq 3N$ and $|E|=e\leq 3N$. Suppose there are $m$ binary equality ($=_2$) in $V_r$. $G'(V'_l\cup V'_r, E')$ is constructed by using a $T\sim(=_2)\sim T^{-1}$ path to replace every $=_2$ in $V_r$. $V'_l=V_l$ and $V'_r$ contains rest vertices in $G'$. $\#G'=\#G$ since $T([1,0,1])T^{-1}=[1,0,1]$.  
		
		\begin{enumerate}[1.]
			\item $\Lambda=\begin{pmatrix} \lambda_1 & 0 \\ 0 & \lambda_2\end{pmatrix}$. 
			If $(\frac{\lambda_1}{\lambda_2})^k=1$, we replace every $=_2$ in $V_r$ by a $2k-1$ length path to obtain $G''(V''_l\cup V''_r, E'')$, in which the vertices are attached with $H$ and $=_2$ interlaced. $\#G''=\#G'=\#G$ and $G''$ is an instance of $\#\{=_1,=_2,=_3\}|\mathscr{F}\cup \{H\}$ with $V''_l=N+(k-1)m\leq (3k-2)N$ and $V''_r= M-m+km\leq 3kN$. Analyzing like Claim 1, Lemma 5.1 is proved. 
			
			\quad\ Otherwise, $=_2$ can be obtained from $\Lambda$ by block interpolation. Dividing $V'_r$ to $m/d$ blocks $B_1,B_2,...,B_{\frac md}$ with each block consisting $d$ binary equality functions. Constructing $G'_{\vec{y}}$ with $\vec{y}=(y_1,y_2,...,y_{\frac md})\in [(d+1)]^{\frac md}$, by replacing every $T(=_2)T^{-1}$ in $B_i$ by a path which is attached with $y_i$ many $H$ (actually it is a $H\sim(=_2)\sim H\sim...\sim H\sim (=_2)\sim H$ path for keeping the bipartite), which is equivalent to $T\Lambda^{y_i} T^{-1}$. $G'_{\vec{y}}$ is an instance of $\#\{=_1,=_2,=_3\}|\mathscr{F}\cup \{H\}$. 
			
			\quad\ Suppose every assignment $S$ of $G'$ has type $t=(t_1,t_2,...,t_{\frac md})^T$, in which $t_i=(t_{i1}, t_{i2})\in \{0,1,...,d\}^2$. 
			$t_{i1}$ records the number of $=_2$ in $B_i$ with two inputs both $0$ and $t_{i2}$ corresponds to the number of $=_2$ which are assigned two $1$ inputs. Then the value of $G_{\vec{y}}$ is:
			\begin{equation}
				\#G_{\vec{y}}=\#G'_{\vec{y}}=\sum_{t} \rho_t \prod_{i=1}^{\frac md} [(\lambda_1)^{t_{i1}}(\lambda_2)^{t_{i2}}]^{y_i}=\sum_{t} \rho_t \prod_{i=1}^{\frac md} (\lambda_2)^{dy_i}(\frac {\lambda_1}{\lambda_2})^{t_{i1}y_i}.
				\label{equ5.1}
			\end{equation} 
			$\rho_t$ is the sum over all type-$t$ satisfied assignments of evaluation on $G'$ with ignoring the binary equality functions in $V'_r$.  
			There are additional condition $t_{i1}+t_{i2}=d$ for all $i$ since only satisfied assignments are considered. So the number of different $t$ are $d+1$. Since $\#G'=\sum_{t}\rho_t$, we can compute it by solving all $\rho_t$.
			
			\quad\ A system of equations can be built after querying $(d+1)$ different $G'_{\vec{y}}$. Its coefficient matrix is tensor product of $A$, which is $(d+1)\times(d+1)$ matrix. $A_{y,t}=(\lambda_2)^{dy}(\frac {\lambda_1}{\lambda_2})^{ty}$ with row indices $y\in\{1,2,...,d+1\}$ and column indices $t\in \{0,1,...,d\}$, so $A$ is the transpose of a Vandermonde matrix. Since $\frac{\lambda_1}{\lambda_2}$ is not root of unity, $A$ is full rank. All $\rho_t$ can be covered by solving the system in $poly(d+1)$ time, then $\#G$ and $\#G'$ both are computed.The total time is $(d+1)\times(poly(m)+T_{oracle}(N_{G'_{\vec{y}}}))+poly(d+1)+poly(d+1)$. The vertices set size of $G'_{\vec{y}}$ are no more than $(N+dm\leq(3d+1)N,M-m+(d+1)m\leq3(d+1)N)$ and edges is less than $e+2dm$.   
			 
			\item $\Lambda=\begin{pmatrix} \lambda & 1 \\ 0 & \lambda \end{pmatrix}$.
			Following case 1, but now every $t_{i1}$ in type $t$ is the number of $=_2$ which are assigned $(0,0)$ or $(1,1)$ and $t_{i2}$ signs how many $=_2$ are assigned $(0,1)$. $\#G=\#G'=\rho_{(\tau,....,\tau)^T}$ where $\tau=(d,0)$ Constructing system of equations with form:
			\begin{equation}
				\#G_{\vec{y}}=\sum_{t} \rho_t \prod_{i=1}^{\frac md} [\lambda^{t_{i1}}({t_{i2}}\lambda^{t_{i2}-1})]^{y_i}=\sum_{t} \rho_t \prod_{i=1}^{\frac md} [{t_{i2}}\lambda^{d-1}]^{y_i},
			\end{equation}
			 since $\Lambda^k=\begin{pmatrix} \lambda^k & k\lambda^{k-1} \\ 0 & \lambda^k \end{pmatrix}$. Now the item of $A$ is $A_{y,t}= t^y \lambda^{(d-1)y}$. The coefficient matrix is still full rank. The system can be solved in $poly(d+1)$ time.
			
		\end{enumerate}
	
		Since $\#\{=_1,=_2,=_3\}|\mathscr{F}\cup \{=_2\}$ is \#ETH-hard, $\#\{=_1,=_2,=_3\}|\mathscr{F}\cup \{H\}$ is also \#ETH-hard by choosing big enough $d$ to establish reduction.
	}  
	~\\
	
	Next considering how $\#\{=_1,=_2,=_3\}|\mathscr{F} \cup {H}$ reduce to $\#\{=_1,=_2,=_3\}|\mathscr{F}$. \cite{complexCSPandR3CSP} has provided the reductions with loosening the restriction of $\mathscr{F}$. We just present the outlines of reducing $\#\{=_1,=_2\}|\mathscr{F} \cup {H}$ to $\#\{=_1,=_2\}|\mathscr{F}$ here, where $\mathscr{F}\not\subseteq\mathscr{D}$ and $H$ is a non-degenerate binary function. 
	
	The first step is apply $M=\frac{1}{\sqrt{2}}\begin{pmatrix}1&1\\1&-1\end{pmatrix}$ to do a local holographic reduction in polynomial time. $\#\{=_1,=_2\}|\mathscr{F} \cup {H} \equiv_T \#\{\delta_0,=_2\}|\mathscr{\tilde{F}} \cup {\tilde{H}}$. $\mathscr{\tilde{F}}=\{M^{\otimes k}F$ | $F\in \mathscr{F}$, $F$ has arity $k$\} is still not the subset of $\mathscr{D}$ and $\tilde{H}=M^{\otimes 2}H$ keeps non-degenerate. 
	
	Secondly, there is always a reduction from $\#\{\delta_0,=_2\}|\mathscr{\tilde{F}} \cup {\tilde{H}}$ to $\#\{\delta_0,=_2\}|\mathscr{\tilde{F}}$ by gadget constructions, if choosing appropriate non-degenerate binary function $\tilde{H}$. For convenience, $\mathscr{\tilde{F}}$ and $\tilde{H}$ are renamed as $\mathscr{F}$ and $H$. The reduction lines are presented in Figure 6.
	
	\begin{figure}[ht]
		\centering
		\includegraphics[scale=0.5]{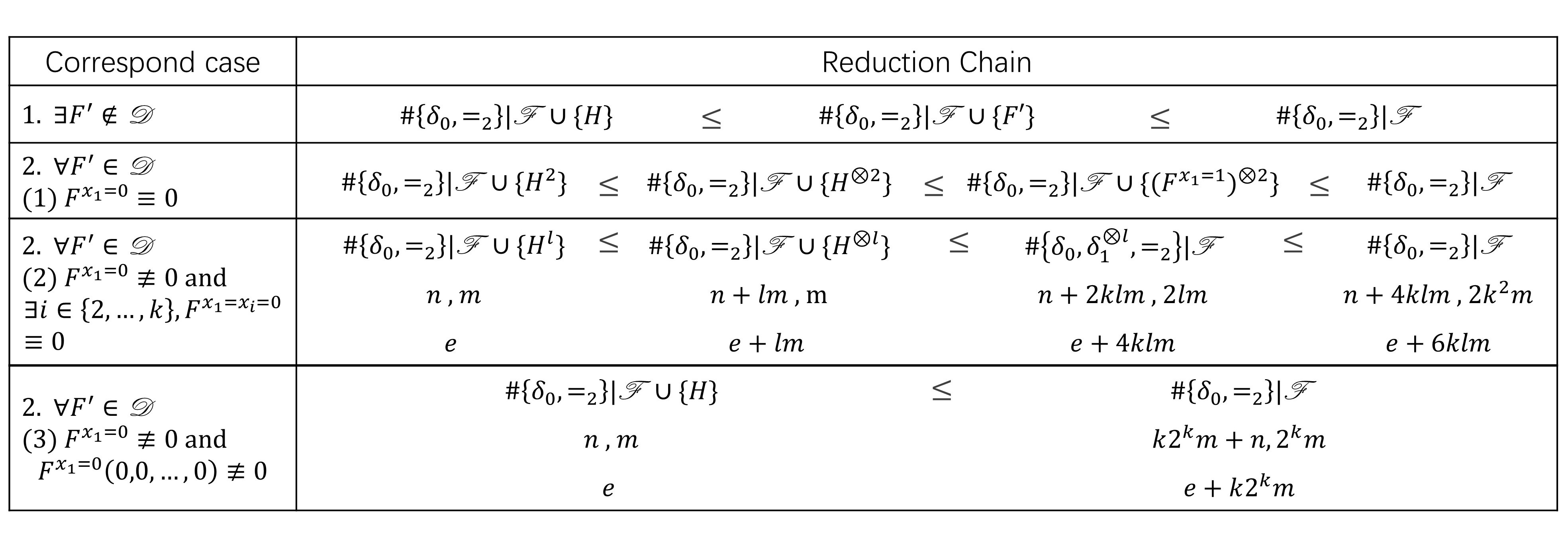}   
		\caption{The upper bounds of corresponding instances' scale in the reduction chains. A triple $(n,m,e)$ is used to represent scale, correspond to the size of $V_l$,$V_r$ and $E$ when $G(V_l\cup V_r,E)$ is an instance of start problem. Since all constructions are constant gadgets, the time is always $poly(m)$ and instances scale are ($O(n),O(n),O(n)$).} 
	\end{figure}
	
	The proof is an induction on the arity of function $F$, $F\in \mathscr{F}-\mathscr{D}$. Either a smaller arity $F'\in \mathscr{F}-\mathscr{D}$ can be realized or a non-degenerate binary function can be directly constructed by gadgets. Paying attention to Case 2-(1) \& (2), $F^{x_1=1}$ and $\delta_1$ can not be constructed directly, but the Kronecker products of them can be obtained. They also work since $H^l$ also are non-degenerate. It can be verified that all the reductions keep \#ETH-hardness transmission. 
	
	We can make a little change to keep every start point is the same problem $\#\{\delta_0,=_2\}|\mathscr{F} \cup {H}$ according Theorem 2.4. This theorem shows such decomposition ($Q^{\otimes l}$ is decomposed to $Q$) still transmits \#ETH-hardness. Hence, we can use $\#\{\delta_0,=_2\}|\mathscr{F} \cup {H}$ to be the start point and all reductions keep \#ETH-hardness spreading, when $\mathscr{F}\notin\mathscr{A}$ and $\mathscr{F}\notin\mathscr{P}$. 
	
	Finally, applying $M^{-1}$ to recover $\mathscr{F}$ and $(=_1)$ by local holographic reduction again ($\#\{\delta_0,=_2\}|\mathscr{\tilde{F}} \equiv_T \#\{=_1,=_2\}|\mathscr{F}$).

	By above three steps and Lemma 5.1, Theorem 1.3 has been proved.

		\section{Conclusion}

	\quad\ In this article, the "FP vs \#P-hard" dichotomy of complex weighted Boolean \#CSP is promoted to "FP vs \#ETH-hard", and the conclusion can even be improved to \#R$_3$-CSP. Besides the dichotomy, an important part in our article is the methods introduced in Section $3$. They all are frequently used in previous researches. I synthesize them and present the internal commonality among them.   
	
	The field of studying the sub-exponential lower bound of counting problem under \#ETH is explored only a tip of the iceberg. There are many open problems such as whether a \#P-hard Holant problem still is \#ETH-hard. And it also is interesting when restrict \#CSP to planar.
		
		\section*{Acknowledge}
		
		\quad\ The author is very grateful to Prof. Mingji Xia for his beneficial guidance and advise.		
		
		\clearpage
		\bibliography{ref}

\begin{thebibliography}{10}

\bibitem{complexHolantc}
Miriam Backens.
\newblock A complete dichotomy for complex-valued holant{\^{}}c.
\newblock In {\em 45th International Colloquium on Automata, Languages, and
  Programming (ICALP 2018)}, volume 107 of {\em LIPIcs}, pages 12:1--12:14.
  Schloss Dagstuhl--Leibniz-Zentrum fuer Informatik, 2018.

\bibitem{ETHCSP}
Cornelius Brand, Holger Dell, and Marc Roth.
\newblock Fine-grained dichotomies for the tutte plane and boolean \#csp.
\newblock {\em Algorithmica}, 81(2):541--556, 2019.

\bibitem{rationalCSP}
Andrei Bulatov, Martin Dyer, Leslie~Ann Goldberg, Markus Jalsenius, and David
  Richerby.
\newblock The complexity of weighted boolean\# csp with mixed signs.
\newblock {\em Theoretical Computer Science}, 410(38-40):3949--3961, 2009.

\bibitem{complexCSP}
Jin-Yi Cai and Xi~Chen.
\newblock Complexity of counting csp with complex weights.
\newblock {\em Journal of the ACM (JACM)}, 64(3):1--39, 2017.

\bibitem{sixvertexmodel}
Jin-Yi Cai, Zhiguo Fu, and Mingji Xia.
\newblock Complexity classification of the six-vertex model.
\newblock {\em Information and Computation}, 259:130--141, 2018.

\bibitem{Holant}
Jin-Yi Cai, Pinyan Lu, and Mingji Xia.
\newblock Holographic algorithms by fibonacci gates and holographic reductions
  for hardness.
\newblock In {\em 49th Annual IEEE Symposium on Foundations of Computer
  Science}, pages 644--653. {IEEE} Computer Society, 2008.

\bibitem{complexCSPandR3CSP}
Jin-Yi Cai, Pinyan Lu, and Mingji Xia.
\newblock The complexity of complex weighted boolean\# csp.
\newblock {\em Journal of Computer and System Sciences}, 80(1):217--236, 2014.

\bibitem{realHolantc}
Jin-Yi Cai, Pinyan Lu, and Mingji Xia.
\newblock Dichotomy for real holant\^c problems.
\newblock In {\em Proceedings of the Twenty-Ninth Annual ACM-SIAM Symposium on
  Discrete Algorithms}, pages 1802--1821. SIAM, 2018.

\bibitem{complexHolant*}
Jin-Yi Cai, Pinyan Lu, and Mingji Xia.
\newblock Dichotomy for holant* problems of boolean domain.
\newblock {\em Theory of Computing Systems}, 64(8):1362--1391, 2020.

\bibitem{ETHquantumGH}
Hubie Chen, Radu Curticapean, and Holger Dell.
\newblock The exponential-time complexity of counting (quantum) graph
  homomorphisms.
\newblock In {\em 45th International Workshop on Graph-Theoretic Concepts in
  Computer Science}, volume 11789, pages 364--378. Springer, 2019.

\bibitem{BooleanCSP}
Nadia Creignou and Miki Hermann.
\newblock Complexity of generalized satisfiability counting problems.
\newblock {\em Information and computation}, 125(1):1--12, 1996.

\bibitem{blockinterpolation}
Radu Curticapean.
\newblock Block interpolation: A framework for tight exponential-time counting
  complexity.
\newblock {\em Information and Computation}, 261:265--280, 2018.

\bibitem{parameterpermanent}
Radu Curticapean and Mingji Xia.
\newblock Parameterizing the permanent: Genus, apices, minors, evaluation mod
  2k.
\newblock In {\em 2015 IEEE 56th Annual Symposium on Foundations of Computer
  Science}, pages 994--1009. IEEE, 2015.

\bibitem{ETHpermTutte}
Holger Dell, Thore Husfeldt, D{\'{a}}niel Marx, Nina Taslaman, and Martin
  Wahlen.
\newblock Exponential time complexity of the permanent and the tutte
  polynomial.
\newblock {\em {ACM} Transaction on Algorithms}, 10(4):21:1--21:32, 2014.

\bibitem{weightedCSP}
Martin Dyer, Leslie~Ann Goldberg, and Mark Jerrum.
\newblock The complexity of weighted boolean\# csp.
\newblock {\em SIAM Journal on Computing}, 38(5):1970--1986, 2009.

\bibitem{GH}
Martin Dyer and Catherine Greenhill.
\newblock The complexity of counting graph homomorphisms.
\newblock {\em Random Structures \& Algorithms}, 17(3-4):260--289, 2000.

\bibitem{ETHKAST}
Russell Impagliazzo and Ramamohan Paturi.
\newblock On the complexity of k-sat.
\newblock {\em Journal of Computer and System Sciences}, 62(2):367--375, 2001.

\bibitem{SETH}
Russell Impagliazzo, Ramamohan Paturi, and Francis Zane.
\newblock Which problems have strongly exponential complexity?
\newblock {\em Journal of Computer and System Sciences}, 63(4):512--530, 2001.

\bibitem{nonnegativeHolant}
Jiabao Lin and Hanpin Wang.
\newblock The complexity of boolean holant problems with nonnegative weights.
\newblock {\em SIAM Journal on Computing}, 47(3):798--828, 2018.

\bibitem{Reflection}
Michael, Freedman, László, Lovász, Alexander, and Schrijver.
\newblock Reflection positivity, rank connectivity, and homomorphism of graphs.
\newblock {\em Journal of the American Mathematical Society}, 20(1):37--51,
  2007.

\bibitem{realHolant}
Shuai Shao and Jin-Yi Cai.
\newblock A dichotomy for real boolean holant problems.
\newblock In {\em 2020 IEEE 61st Annual Symposium on Foundations of Computer
  Science}, pages 1091--1102, 2020.

\bibitem{phardproblems}
Salil~P. Vadhan.
\newblock The complexity of counting in sparse, regular, and planar graphs.
\newblock {\em {SIAM} Journal on Computing}, 31(2):398--427, 2001.

\bibitem{CCP}
Leslie~G Valiant.
\newblock The complexity of computing the permanent.
\newblock {\em Theoretical computer science}, 8(2):189--201, 1979.

\bibitem{completeproblems}
Leslie~G Valiant.
\newblock The complexity of enumeration and reliability problems.
\newblock {\em SIAM Journal on Computing}, 8(3):410--421, 1979.

\bibitem{ValiantAccidental}
Leslie~G Valiant.
\newblock Accidental algorthims.
\newblock In {\em 47th Annual IEEE Symposium on Foundations of Computer
  Science}, 2006.

\bibitem{LeslieHolographic}
Leslie~G Valiant.
\newblock Holographic algorithms.
\newblock {\em SIAM Journal on Computing}, 37(5):1565–1594, 2008.

\end{thebibliography}
		
	\end{spacing}

\end{document}